\documentclass[twocolumn]{aastex631} 

\usepackage{graphicx}
\usepackage{xspace}
\usepackage{tabularx}
\usepackage{enumitem}
\usepackage{nicefrac}
\usepackage{soul}
\usepackage{appendix}
\usepackage{amssymb}
\usepackage{soul}
\usepackage{amsmath}
\usepackage{lipsum, babel}

\newcommand{\gyro}{gyrochronology\xspace}

\newcommand{\total}{765\xspace}  
\newcommand{\newdet}{50\xspace} 
\newcommand{\olddet}{715\xspace} 
\newcommand{\sample}{4346\xspace}  
\newcommand{\planets}{101\xspace}  
\newcommand{\withprot}{451\xspace}  

\newcommand{\dnu}{$\Delta \nu$\xspace}
\newcommand{\rhk}{$\log{R'_{\text{HK}}}$\xspace}
\newcommand{\numax}{$\nu_{\text{max}}$\xspace}
\newcommand{\muhz}{$\mu \text{Hz}$\xspace}
\newcommand{\teff}{T$_{\text{eff}}$\xspace}
\newcommand{\logg}{$\log g$\xspace}
\newcommand{\prot}{P$_\text{rot}$\xspace}
\newcommand{\kp}{$Kp$\xspace}

\newcommand{\solrad}{R$_\odot$\xspace}
\newcommand{\rearth}{R$_\oplus$\xspace}
\newcommand{\app}{$\sim$}
\newcommand{\snr}{signal--to--noise\xspace}

\newcommand{\kepler}{\textit{Kepler}\xspace}
\newcommand{\tess}{TESS\xspace}
\newcommand{\gaia}{\textit{Gaia}\xspace}
\newcommand{\syd}{\texttt{SYD}\xspace}
\newcommand{\pysyd}{\texttt{pySYD}\xspace}
\newcommand{\lkurve}{\texttt{lightkurve}\xspace}
\newcommand{\hires}{{\small HIRES}\xspace}

\newcommand{\serenelli}{\cite{Serenelli_2017}}
\newcommand{\chaplin}{\cite{Chaplin_2014}}
\newcommand{\balona}{\cite{Balona_2020}}
\newcommand{\mathur}{\cite{mathur_2021}}

\newcommand{\new}[1]{\textcolor{blue}{#1}}

\shorttitle{Solar--Like Oscillators in Kepler Short--Cadence Data}

\graphicspath{{./}{figures/}}
\begin{document}

\title{A Homogeneous Catalog of Oscillating Solar--Type Stars Observed by the Kepler Mission and a New Amplitude Scaling Relation Including Chromospheric Activity}

\correspondingauthor{Maryum Sayeed}
\email{maryum.sayeed@columbia.edu}

\author[0000-0001-6180-8482]{Maryum Sayeed}
\affiliation{Department of Astronomy, Columbia University, 550 West 120th Street, New York, NY 10027, USA}

\author[0000-0001-8832-4488]{Daniel Huber}
\affiliation{Institute for Astronomy, University of Hawai`i, 2680 Woodlawn Drive, Honolulu, HI 96822, USA}
\affiliation{Sydney Institute for Astronomy (SIfA), School of Physics, University of Sydney, NSW 2006, Australia}

\author[0000-0003-1125-2564]{Ashley Chontos}
\affiliation{Department of Astrophysical Sciences, Princeton University, 4 Ivy Lane, Princeton, NJ 08544, USA}

\author[0000-0003-3020-4437]{Yaguang Li}
\affiliation{Institute for Astronomy, University of Hawai`i, 2680 Woodlawn Drive, Honolulu, HI 96822, USA}

\begin{abstract}
We present a homogeneous catalog of global asteroseismic parameters and derived stellar parameters for \total \kepler main--sequence and subgiant stars. The catalog was produced by re--analyzing all available \kepler DR25 short--cadence data using \pysyd, an automated pipeline to extract global asteroseismic parameters. We find \newdet new detections, seven of which are also planet candidate host stars. We find excellent agreement between our \numax and \dnu\ measurements and literature values, with an average offset of $0.2 \pm 0.4\%$ ($\sigma=5$\%) and $0.2 \pm 0.7\%$ ($\sigma=2$\%), respectively. In addition, we derive stellar radii and masses with an average precision of 2.7\% and 10.4\%, respectively, and find a mean offset of $0.8 \pm 0.2$\% ($\sigma=6$\%) between our radii derived with asteroseismology and those from \gaia parallaxes. Using spectroscopic \rhk activity measurements from Keck/\hires, we derive an amplitude scaling relation with an activity term for main--sequence and subgiant stars, which successfully predicts amplitudes with a precision of $\approx$\,8--9\%. 
Our work is the largest and most homogeneous asteroseismic catalog of \kepler main--sequence and subgiant stars to date, including a total of \planets stars hosting planet candidates and \withprot stars with measured rotation periods.
\end{abstract}
\keywords{Asteroseismology (73) -- fundamental parameters (555) -- stellar oscillations (1617) -- photometry (1234)}

\section{Introduction} \label{sec:intro}

Asteroseismology, the study of stellar oscillations, is a powerful method for constraining fundamental stellar properties. Advances in photometric time--series observations from space--based missions like CoRoT \citep{corot_baglin}, \kepler \citep{kepler_borucki, kepler_koch} and NASA's Transiting Exoplanet Survey Satellite \citep[TESS,][]{tess_ricker} have revolutionized asteroseismology for stars across the Hertzsprung--Russell (H--R) diagram. This technique provides valuable insight into stellar interiors including classification of evolutionary stages \citep[e.g.][]{bedding_2011}, measurement of internal rotation \citep[e.g.,][]{beck_2012}, and detection of magnetic fields \citep[e.g.,][]{Fuller_2015, cantiello_2016, Stello2016, li_2020}. Furthermore, it advances our knowledge of galaxy evolution \citep[e.g.,][]{hon_2021}, and enables precise characterization of exoplanet host stars \citep[e.g.,][]{johnson_2010, Huber_2013, Lundkvist2016}. 

For solar--like oscillations, the power excess can be characterized through two global asteroseismic parameters: the frequency of maximum power (\numax) and the average large frequency separation (\dnu). Combined with scaling relations, these parameters provide an efficient method to derive fundamental stellar parameters such as mass, radius, surface gravity, and density. 
More precise stellar properties such as ages can be inferred through modeling individual frequencies \citep[e.g.,][]{Metcalfe_2010, batalha_2011, howell_2012, carter_2012, Metcalfe_2012, silva_2013, gilliland_2013, chaplin_2013, silva_2015, silva_2017}. However, this is often not possible for stars with low \snr (SNR) or short time--series with limited frequency resolution. Large, homogeneous catalogs of global asteroseismic parameters therefore remain a fundamental benchmark in stellar astrophysics.

In addition to inferring ages, large catalogs of asteroseismic parameters enable the study of the effect of stellar activity on suppressed solar--like oscillations. Stellar activity and magnetic fields are known to decrease oscillation amplitudes \citep{chaplin_2000, komm_2000, Garcia2010, Mathur2019}, but this effect has neither been quantified nor incorporated in amplitude scaling relations. The mechanism for this amplitude suppression is still unclear, but stellar activity has been suggested as a dominating factor \citep[e.g.,][]{Mosser2009, Dall2010, gaulme2014, Corsaro2024}. For instance, some have studied the effects of binary interactions \citep{Gehan2022, Gehan2024} and chromospheric activity \citep[e.g.,][]{Bonanno2014} on oscillation amplitudes. Large, homogeneous catalogs of solar--like oscillators which overlap with spectroscopic data are required to advance our understanding of amplitude scaling relations.

NASA's \kepler\ mission observed \app196,000 stars in long--cadence (29.4 minutes) mode, which is sufficient to sample oscillations in red giant stars \citep[e.g.,][]{hekker_2011b, Huber_2011, stello_2013, huber_2014, mathur_2016, yu_2016}. Main--sequence and subgiant stars require sampling in short--cadence (58.5 seconds), which was obtained for only \app500 pre--selected stars during each observing quarter; 
fundamental parameters for this sample were provided in \cite{Chaplin_2014}. Since then, the \kepler Science Office released newly calibrated data with improved data processing, such as smear correction and aperture image extension \citep[][]{thompson_2016}. Furthermore, while there have been additional discoveries of seismic detections in main--sequence and subgiant stars  \citep[e.g.,][]{white_2012, creevey_2012, gaulme_2013, dogan_2013, yaguang_2020, Balona_2020, mathur_2021, Bhalotia2024}, they employ different methods to measure global asteroseismic quantities, which can lead to systematic offsets in derived stellar properties across the H--R diagram.


In this paper, we use a single, well--tested and open--source asteroseismic pipeline \citep[\pysyd,][]{chontos_2022}, to extract global asteroseismic parameters and determine stellar properties for main--sequence and subgiant stars observed in \kepler re--calibrated short--cadence data. \pysyd is adapted from the framework of the IDL--based \syd pipeline, and has been used to extract asteroseismic parameters for many \kepler stars \citep[e.g.,][]{Huber_2011, bastien_2013, Chaplin_2014, Serenelli_2017, Yu_2018}. We provide a homogeneous catalog of asteroseismic stellar masses, radii, and global oscillation parameters for \total solar--like oscillators in \kepler.

\begin{deluxetable}{l|ccc}[t!]
\tabletypesize{\footnotesize}
    \tablewidth{0pt}
    \tablecaption{Summary of previously known solar--like oscillators re--analyzed with \pysyd for this work. Section \ref{sec:target} provides reasoning for the difference in the number of stars in a given catalog and those re--analyzed. \label{tb:tb-sample}}
    \tablehead{ & \multicolumn{3}{c|}{Targets}  \\
    \multicolumn{1}{c|}{Source catalog} & \multicolumn{1}{c}{Total} &\multicolumn{1}{c}{Re--analyzed} & \multicolumn{1}{c|}{Confirmed} }
    \startdata
    \cite{Serenelli_2017} & 415 &398 & 396 \\
    \cite{Chaplin_2014} &518& 111 & 111 \\
    \cite{mathur_2021} &624 & 97 & 94 \\
    \cite{Balona_2020} & 70& 13 & 11 \\
    \cite{Huber_2013} & 77 & 72 & 70 \\
    \cite{Lundkvist2016} & 102&38 & 25 \\
    \cite{li_2020} & 36 &1 &1 \\
    \cite{Pinsonneault_2018} & 6676 &1 & 1 \\
    \cite{Lund_2017} & 66 &2 & 2\\
    \cite{mosser_2014} & 1178 &1 & 1\\
    \cite{white_2012} & 163 & 1 & 1  \\ 
    \cite{Chontos_2019} & 1 & 1 & 1 \\
    \cite{Bhalotia2024} & 1 & 1 & 1 \\
    \enddata
\end{deluxetable}

\section{Observations and Data Analysis} \label{sec:style}

\subsection{Target Selection}\label{sec:target}
We started with the full list of \kepler stars observed in short--cadence from Mikulski Archive for Space Telescopes (MAST), resulting in a total of 5678 unique stars. We first removed stars with detected oscillations in long--cadence data \citep{Yu_2018} (5391 remaining), eclipsing binaries \citep{kirk_2016} (4810 remaining), and Delta Scuti pulsators \citep{murphy_2019} which are too hot to show solar--like oscillations. For the remaining \sample targets, we flagged stars that were already reported in previous studies to keep track of new detections in our sample. The following is a breakdown of previous asteroseismic detections:
\begin{enumerate}[label=(\roman*)]
    \setlength\itemsep{0em}
    \setlength\parskip{0em}
    \item 415 stars from \cite{Serenelli_2017}, but one was a binary \citep{kirk_2016}, two were observed in long--cadence \citep{Yu_2018}, and 14 are exoplanet hosts; 398 stars remain in which to search for solar--like oscillations.
    \item 518 stars from \cite{Chaplin_2014}, but 407 overlapped with S17; 111 stars remain.
    \item 624 stars from \cite{mathur_2021}, but 524 were already known detections, and three were found to be binaries; 97 stars remain.
    \item 70 stars from \cite{Balona_2020}, but 56 were already known detections (in \cite{Serenelli_2017}, \cite{Chaplin_2014}, or \cite{mathur_2021}), and one was an eclipsing binary; 13 stars remain.
    \item 117 exoplanet hosts from \cite{Huber_2013} and \cite{Lundkvist2016}, but three were not observed in short--cadence (KIC 4476423, 8219268, 9088780), and six are binaries (KIC 2306756, 4769799, 5652983, 6678383, 8554498, 8803882); 108 stars remain. 
\end{enumerate}

\begin{figure*}[t!]
\begin{center}
\includegraphics[width=\linewidth,angle=0]{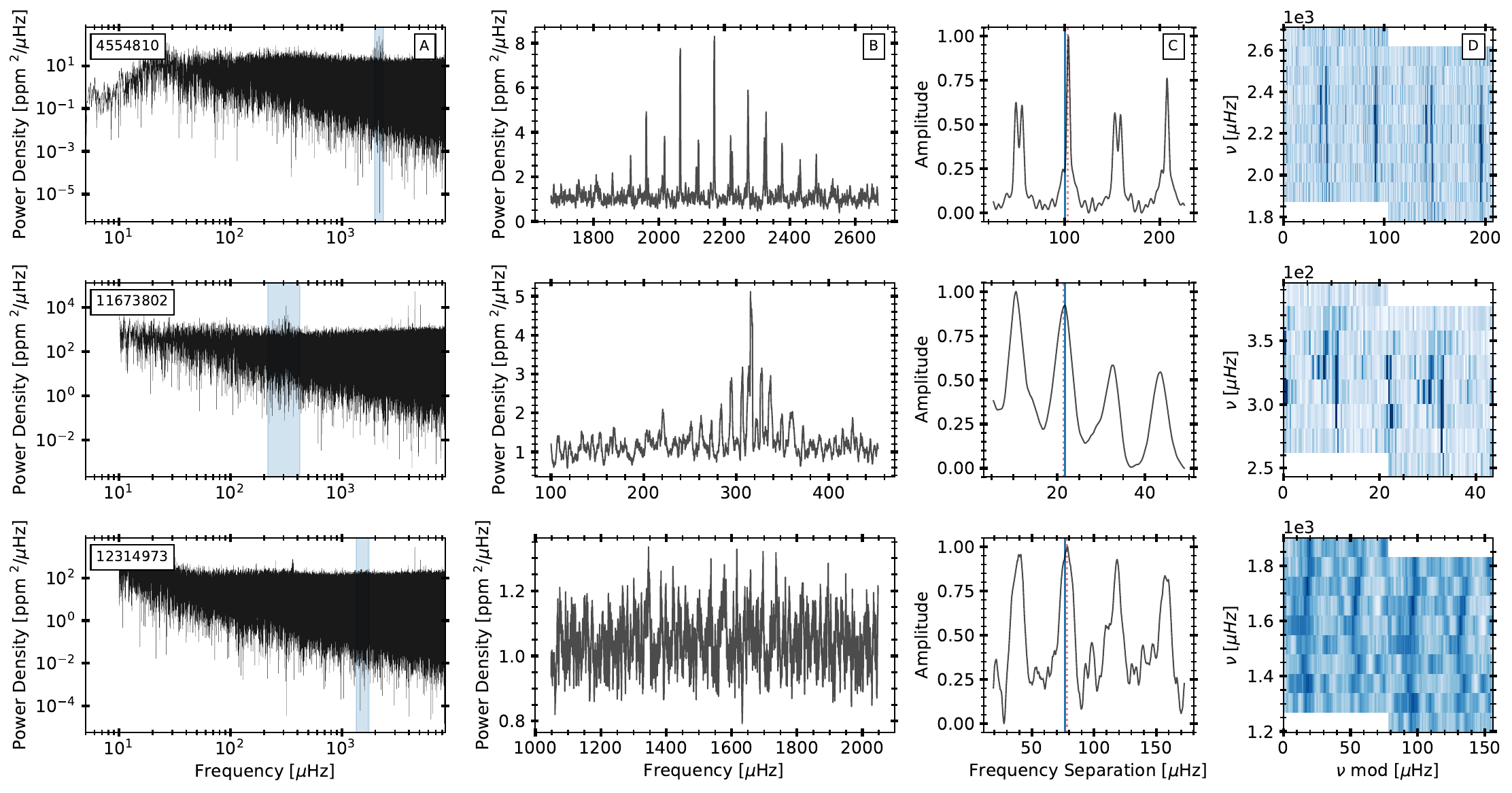}
\caption{Example data for three \kepler stars with high SNR (top panels, KIC\, ID 4554810), moderate SNR (middle panels, KIC\, ID 11673802), and low SNR (bottom panels, KIC\, ID 12314973). \textit{Column A:} Critically sampled power density spectrum (in units of ppm$^2/\mu$Hz). The blue band is centered on the \pysyd's estimated \numax value. \textit{Column B:} Background corrected power density spectrum in linear scale centered on \pysyd's estimate of \numax. \textit{Column C:} Auto--correlation function of the data shown in column B. The blue line corresponds to the expected \dnu value given the input \numax, and the red dotted line corresponds to the measured \dnu using \pysyd's analysis. \textit{Column D:} \'Echelle diagram of the background corrected power spectrum (shown in column B) using \pysyd's measurement of \dnu.}
\label{fig:data}
\end{center}
\end{figure*}

We also performed a systematic search for oscillations in \kepler exoplanet hosts. Of the total 3611 stars given an exoplanet disposition of confirmed or candidate on the NASA Exoplanet Archive \citep{koidr25}, 766 stars were observed in short--cadence, of which eight targets were removed since they are classified as Delta Scuti pulsators or already detections in long--cadence data \citep{Yu_2018}. In summary, we re--analyzed \kepler DR25 data for 4346 stars, of which 758 were confirmed or candidate exoplanet hosts. Table \ref{tb:tb-sample} provides a breakdown of the number of targets analyzed from each source catalog. 




\subsection{Data Preparation}\label{sec:data}

Before running our analysis, we prepared all light curves using the following steps: 

\begin{enumerate}[label=(\roman*)]
    \setlength\itemsep{0em}
    \setlength\parskip{0em}
    \item We downloaded Pre--Search Data Conditioning Simple Aperture Photometry (PDCSAP) from MAST using \lkurve \citep{lkurve} with good quality flags (\texttt{SAP\_QUALITY=0)}. All available \kepler quarters were stitched together using the \texttt{stitch()} function in \lkurve, and used in the subsequent analysis.
    \item We performed a sigma--clipping routine, and removed outliers greater than 5$\sigma$. 
    \item We applied a smoothing Savitzky--Golay filter of 1--day to remove any long--periodic (low--frequency) variations, and normalized the resulting flux. 
    \item We calculated a critically sampled power density spectrum for each star (ie. where the frequency resolution is inverse of the total duration of the time series data) which was fed into the detection pipeline.
\end{enumerate}

For some targets, there were exceptions in how the data were processed. These targets and exceptions are described below:

\begin{enumerate}[label=(\roman*)]
    \setlength\itemsep{0em}
    \setlength\parskip{0em}
    \item For some \kepler targets that were previous detections, DR25 did not improve the data quality and therefore did not reveal oscillations using \pysyd. Therefore for a subset of the sample, we used the same data as \cite{Chaplin_2014}. These stars are KIC IDs 2998253,  3437637,  3547794, 4465324, 4646780, 5265656,  5689219,  6034893, 6853020,  7465072, 8360349,  8656342, 10130724, 11802968, 11862119.
    
    \item For KIC IDs 6278762 and 7051180, we used pre--prepared power spectra from KASOC\footnote{\textit{Kepler} Asteroseismic Science Operations Center} \new{\citep{handberglund2014}} instead of available DR25 data from \lkurve due to higher noise in the data. For instance, the white noise $\nu=6000-8496 \, \mu {\rm Hz}$ \muhz is $0.6 \; \text{ppm}^2 /\mu \text{Hz}$ in the KASOC data, but $51 \; \text{ppm}^2 /\mu \text{Hz}$ in DR25 data for KIC ID 6278762. For KIC ID 7051180, the white noise is $41 \; \text{ppm}^2 /\mu \text{Hz}$ and $80 \; \text{ppm}^2 /\mu \text{Hz}$, respectively. 
    \item For seven stars, we performed notching (ie. masking out the $l=1$ modes) to remove mixed modes in order to measure \dnu confidently. These stars are primarily subgiants, KIC IDs 5683538, 5939450, 7669332, 9664694, 9894195, 10593351, 11397541.
    \item For KIC ID 8360349, we used only the first month of data since the photometry in the later three months was noisier. For instance, the flux scatter in the first month was \app$140$ ppm, an order of magnitude smaller than the remaining three months: \app$1900$ ppm, $3200$ ppm, $1900$ ppm, respectively. Similarly, for KIC ID 6106120, we used the first and tenth month of data.
    \item For two planet hosts in \cite{Huber_2013} -- KIC IDs 4141376 and 5514383 -- we did not find oscillations with DR25 data, and therefore used data from the original paper in \cite{Huber_2013}.
    \item For KIC ID 3861595, we used the power spectrum from the original discovery paper, \cite{Chontos_2019}, to confirm the detection.
\end{enumerate}

\subsection{Detection Pipeline} \label{sec:pysyd}

To detect solar--like oscillations and measure global asteroseismic parameters, we used \pysyd, an open-source Python package \citep{chontos_2022}, adapted from the IDL--based \syd pipeline \citep{Huber_IDL}. While \pysyd uses the same framework and steps as \syd, it has improved features such as automated background model selection, estimation of the large frequency separation with Gaussian weighting, and automated white noise calculation \citep{chontos_2022}. We briefly describe the basic analysis steps of \pysyd below, and refer the reader to \cite{chontos_2022} and \cite{Huber_2011} (hereafter H11) for a more detailed discussion. 

\pysyd consists of three primary steps: (a) automatic identification of the power excess due to solar--like oscillations, (b) optimization and correction for the stellar background contribution, and (c) calculation of global asteroseismic parameters: the mean large frequency spacing (\dnu), and frequency of maximum oscillation (\numax). The pipeline first locates the region of power excess in step (a) in order to exclude this region during the background modeling and correction performed in step (b). The background, generated by granulation and stellar activity, is modelled by a sum of power laws \citep{Harvey_1985}. The frequency at which the maximum amplitude occurs in the smoothed power spectrum is taken as the frequency of maximum oscillation, \numax \citep{Kjeldsen_2008}. The fitted background model is then subtracted from the power spectrum in step (b), and the residual power spectrum is used to calculate the autocorrelation function (ACF) collapsed over all frequency spacings in step (c). The highest peaks in the ACF are found, and the peak closest to the expected \dnu is considered the best estimate of the given target's \dnu. Uncertainties in global asteroseismic values are estimated with Monte--Carlo simulations. Figure \ref{fig:data} shows data generated by \pysyd for three stars in our sample with varying \snr. 



\subsection{New Detections and Comparison with Previous Results}
\begin{figure}[t!]
\begin{center}
\includegraphics[width=\linewidth,angle=0]{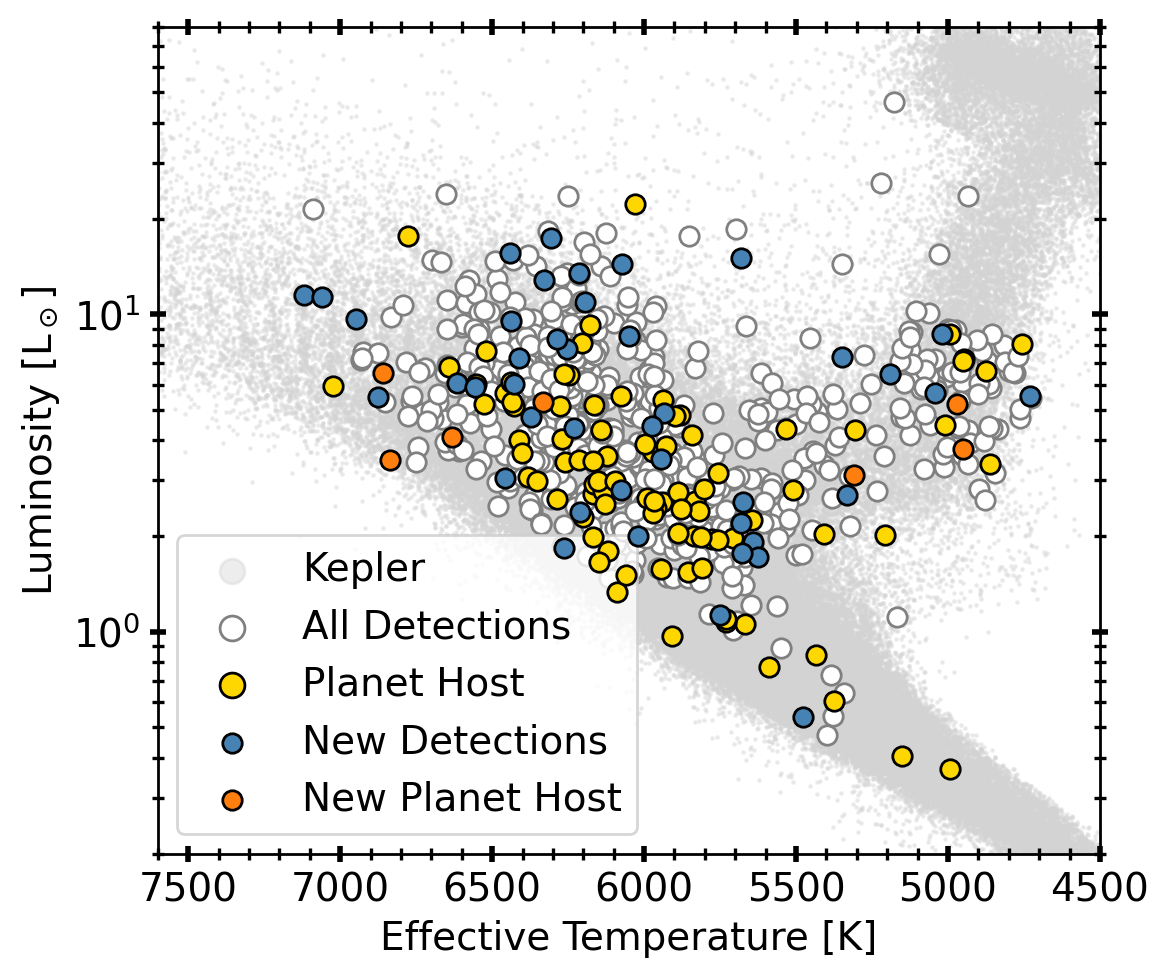}
\caption{H--R diagram showing \total stars with detected oscillations in white circles, new detections in blue, planet hosts in yellow, and new planet hosts in orange. The complete \kepler sample is shown in grey for reference, using stellar parameters from \cite{berger_2020}.}
\label{fig:hrd}
\end{center}
\end{figure}

We detected solar--like oscillations in \total \kepler stars observed in short--cadence, where \newdet stars are new detections and \olddet are already known detections. Figures \ref{fig:new_dets_1} and \ref{fig:new_dets_2} in the Appendix show the smoothed, background corrected spectrum for the \newdet new detections centered on \numax. For all detections, Table \ref{tb:pysyd-output} provides the \kepler Input Catalog (KIC) ID, source flag to indicate the original source of asteroseismic measurements, planet flag to indicate whether the star is a planet host, and seismic measurements, and Table \ref{tb:misc} provides relevant stellar parameters for all sources. The \pysyd output was confirmed through visual inspection, requiring a clear power excess (using an estimated \numax derived with scaling relation and stellar parameters from \cite{berger_2020}), evidence of vertical ridges in the \'echelle diagram, and periodic signal in the ACF. For 34 stars, the \snr is too low to report a reliable measurement of \numax from a heavily smoothed power spectrum, consistent with previous catalogs in which only \dnu\ measurements were feasible for main--sequence and subgiant oscillators \citep{Chaplin_2014}.  Similarly, only the \numax measurement is provided for 13 stars when \dnu cannot be confidently measured. Figure \ref{fig:hrd} shows the complete sample on a H--R diagram using stellar parameters from \cite{berger_2020}, derived with \gaia parallaxes. Our new detections are randomly distributed among the previously known sample; this is consistent with most new detections coming from the improved photometric precision of reprocessed short--cadence light curves. 

\begin{figure*}[t!]
\begin{center}
\includegraphics[width=\linewidth,angle=0]{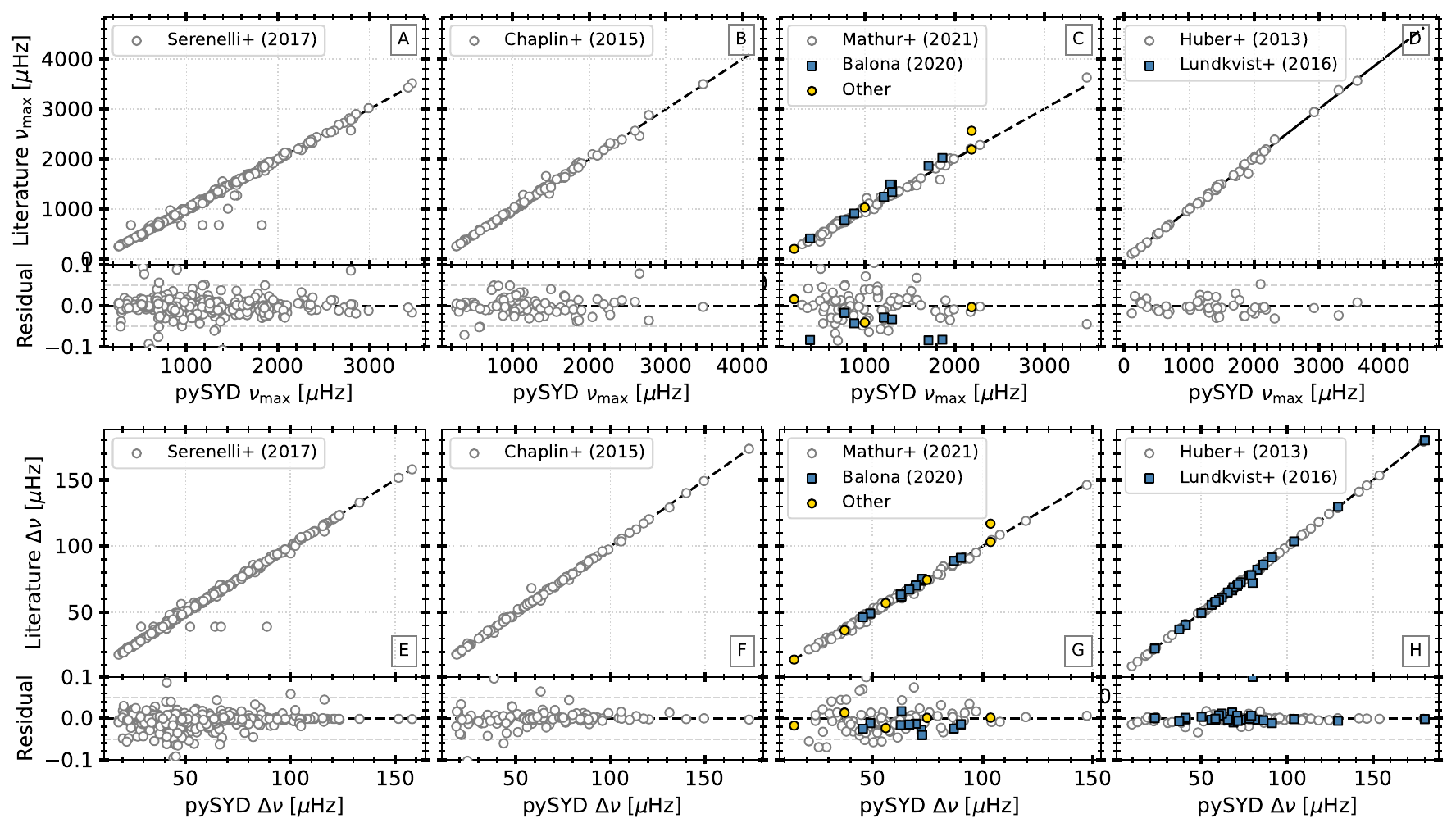}
\caption{Comparison of \numax (top row) and \dnu (bottom panel) measurements from this work and previous catalogs, including \serenelli, \chaplin, \mathur, \balona, \cite{Huber_2013}, \cite{Lundkvist2016}, and others (summarized in Section \ref{sec:target} and Table \ref{tb:pysyd-output}). The bottom panel of each subplot shows the residuals.}
\label{fig:lit_comp}
\end{center}
\end{figure*}

We compared our global asteroseismic parameters from \pysyd to those in literature. Most targets were found in \cite{Serenelli_2017} and \cite{Chaplin_2014}, while other detections were found in separate studies including  \cite{white_2012} (KIC ID 11290197), \cite{mosser_2014} (KIC ID 6776673), \cite{Lund_2017} (KIC IDs 12069424 and 12069449), \cite{Pinsonneault_2018} (KIC ID 3831992), and \cite{li_2020} (KIC ID 10920273). Published values were obtained from the most recent study; however, if a star existed in \cite{Chaplin_2014} as well as a more recent study, measurements from the former were prioritized. Similarly, exoplanet hosts with detections were published in \cite{Huber_2013} (hereafter H13) and \cite{Lundkvist2016} (hereafter L16), but measurements for 40 detections were available in both catalogs; for overlapping stars, we compared our measurements to those in H13 given the similarity of the pipeline in H13 and this work. Therefore, of the 114 planet hosts observed in short--cadence, we re--analyzed all 71 in H13 and 38 in L16. Table \ref{tb:tb-sample} summarizes the number of stars in each catalog. See Section \ref{sec:target} for the reasoning behind the difference in the number of stars in a given catalog and those re--analyzed.

All targets in \cite{Chaplin_2014} that were re--analyzed were confirmed as detections in our analysis. Two targets in \cite{Serenelli_2017} (KIC IDs 3730801, 11075448), three in \cite{mathur_2021} (KIC IDs 7418476, 9109988, 10969935), and two in \cite{Balona_2020} (KIC IDs 6048403, 7833587) were not confirmed as detections. Three of these four stars had only one month of short--cadence data, which could be responsible for low data quality and difficulty in detecting oscillations.

In the sample of planet hosts from H13, we did not recover oscillations in four stars using DR25 data, but we detected oscillations in two when using data from the original paper: KIC IDs 4141376 and 5514383. The third star, KIC ID 6032981, has only one month of short--cadence data and H13 used long--cadence data to detect oscillations. For the fourth star, KIC ID 10593626, we did not recover oscillations in the DR25 data and we do not have access to the data from the original paper, \cite{kepler_borucki}. Similarly, there were 13 stars in L16 for which we did not find oscillations using DR25 data nor the data from the original paper: KIC IDs 4815520, 5383248, 6678383, 7887791, 7941200, 8753657, 9072639, 9579641, 10026544, 10130039, 10748390, 11600889, 11623629. Only seven had a greater than 80\% detection probability while the remaining six had less than 50\% detection probability. Figure \ref{fig:unconfirmed_dets} in the Appendix shows the power spectra of the 22 previously known detections for which we did not find solar--like oscillations. Table \ref{tb:unconfirmed} includes the KIC IDs, source flag, and measurements of \numax and \dnu of the 22 targets which we did not confirm as detections.

Figure \ref{fig:lit_comp} compares our measurements of \numax and \dnu with previously measured values. We observe excellent agreement between our values and previously measured values, with an average offset of $0.2 \pm 0.7 \%$ and scatter of 5.3\% in \numax, and $0.2 \pm 0.4 \%$ and scatter of 2.5\% in \dnu, where the scatter is the robust standard deviation of the residuals and the offset is the mean of the residuals.
We further investigated outliers\footnote{KIC IDs 4574610, 4914923, 6442183, 7341231, 7747078, 11070918, 12366681} in \numax comparison between our measurements and those from \cite{Serenelli_2017} (Figure \ref{fig:lit_comp}a). There are seven\footnote{KIC IDs 4641621, 5344612, 5431016, 7611858, 7910848, 8656342} stars with identical values of \numax in \cite{Serenelli_2017} (\numax $=690.176$ \muhz) producing a collection of points horizontally, and resulting in a large deviation between our values and those published. Similarly, the outliers in Panel E of Figure \ref{fig:lit_comp} all have the same values of \numax and \dnu in \cite{Serenelli_2017}. In addition to these seven targets, there are additional six stars with a fraction difference greater than 10\% between our measurement and those from \cite{Serenelli_2017}. Five of the six targets have one month of data, but the sixth target -- KIC ID 5431016 -- is a $F-$type star with 16 months of data. $F-$type stars show increased line widths causing the $l=0$ and $l=2$ modes to blend together, making it difficult to distinguish between $l=1$ and $l=0,2$ modes \citep[e.g.,][]{white_2012}; therefore the discrepancy between the two measurements could be caused by the broad shape of the Gaussian envelope.

\subsection{Spectroscopic Activity Indicators}
Stellar chromospheric activity shows indicators in the line cores of Ca {\small II} H and K, Mg {\small II}, H$\alpha$ and Ca infrared triplet \citep[e.g.,][]{Hall2008, Stassun2012}. The two most popular activity indices are $S_{\rm HK}$, defined as the ratio of flux in Ca {\small II} H and K lines to flux in nearby continuum region, and \rhk, the ratio between chromospheric flux and bolometric flux after removal of photospheric flux, which is non--negligible for solar--like stars \citep[e.g.,][]{Hartmann1984, Noyes1984}. Compared to $S_{\rm HK}$, \rhk is more appropriate for comparing Ca {\small II} H and K emission on the same scale for stars of multiple spectral types \citep[e.g.,][]{borosaikia_2018}.  

We derive $S_{\rm HK}$ values from publicly available spectra obtained with Keck/\hires \citep{Isaacson2024} using the method described in \cite{Isaacson10}. The spectra have an average SNR of 85 at $\approx$\,550\,nm and were in part obtained to calibrate the performance of spectroscopic analysis pipelines to derive fundamental stellar parameters \citep{furlan18}. We perform quality cuts to ensure reliable $S_{\rm HK}$ measurements, using only $S_{\rm HK}$ $>0.10$ and SNR $> 8$. To calculate \rhk, we require both $S_{\rm HK}$ and $B-V$ which can be derived using input \teff, \logg, and [Fe/H]. Although there exist methods to derive \rhk directly from $S_{\rm HK}$ without using \teff, \logg, and [Fe/H], we followed the method from \cite{Noyes1984} given its applicability across spectral types. For instance, the method described in \cite{Lorenzo-Oliveira_2018} was calibrated on solar twins, while the one in \cite{Marvin2023} results in a 0.2 dex offset compared to \cite{Noyes1984} at low activity. We use the following relation and coefficients from \cite{Sekiguchi2000} to derive $B-V$,
\begin{equation}
\begin{split}
    (B-V)_0 =& \; t_0 + t_1\log(T_{\rm eff}) + t_2\log(T_{\rm eff})^2 \\ 
    & + t_3\log(T_{\rm eff})^3 + f_1[{\rm Fe/H}] + f_2[{\rm Fe/H}]^2 \\
    & + d_1[{\rm Fe/H}]\log(T_{\rm eff}) + g_1\log(g) \\
    & + e_1\log(g) \log(T_{\rm eff})
    \label{eq:b-v}
\end{split}
\end{equation}
where we use \teff, \logg, and [Fe/H] from \texttt{Specmatch-synth}  \citep{specmatch}, a tool to extract fundamental atmospheric parameters by fitting synthetic model atmospheres from \cite{Coelho2005} to optical spectra.

We then use $S_{\rm HK}$ (on the Mount Wilson scale) and $B-V$ to derive \rhk using the updated relation from \cite{Noyes1984}, adapted from \cite{Middelkoop1982},

\begin{equation}
    R_{\rm HK} = 1.340 \times 10^{-4} \; C_{cf} \;S_{\rm HK} 
\end{equation}
where $\log C_{cf} = 1.13(B-V)^3 - 3.91(B-V)^2+2.84(B-V)-0.47$. However, \cite{Noyes1984} corrected $C_{cf}$ for a non--physical maximum at $B-V = 0.43$, so that $\log C'_{cf} = \log C_{cf} + \Delta \log C$, where $\Delta \log C = 0$ if $B-V > 0.63$. Otherwise, $\Delta \log C $ is described by the following,

\begin{equation}
\begin{split}
    \Delta \log C = \; & 0.135x - 0.814x^2 + 6.03x^3, \\ 
    {\rm where} \; & x =  0.63 - (B-V)
\end{split}
\end{equation}

Finally, to derive \rhk from $R_{\rm HK}$, 
\begin{equation}
\begin{split}
    \log{R'_{\rm HK}} = & \log{R_{\rm HK}} - \log{R_{\rm PHOT}},\\
    \log{R_{\rm PHOT}} = & -4.898 + 1.918(B-V)^2\\
                        & -2.893(B-V)^3
\end{split}
\end{equation}

\begin{figure}[t!]
\begin{center}
\includegraphics[width=\linewidth,angle=0]{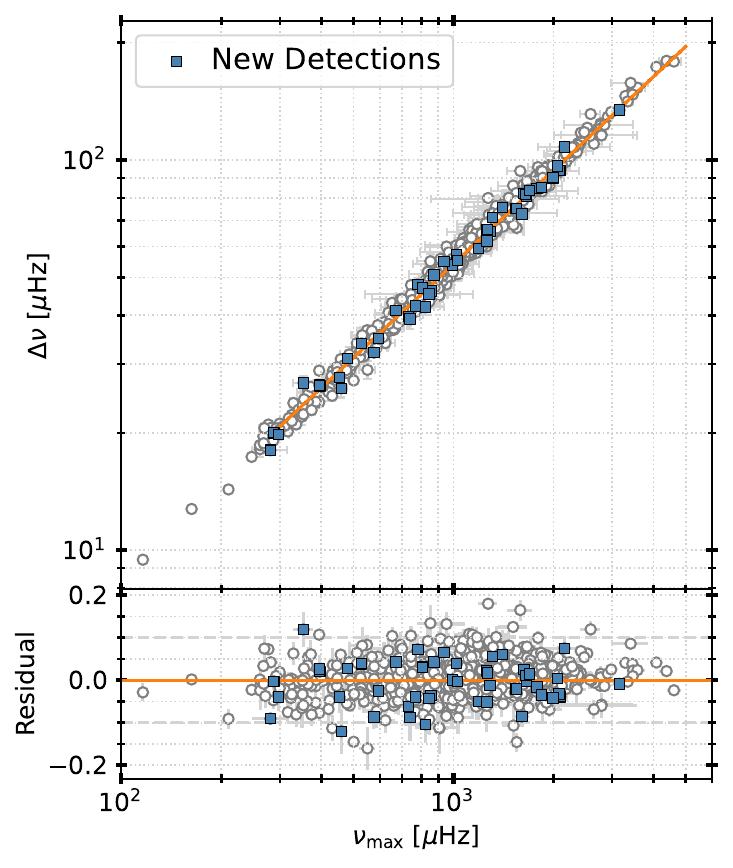}
\caption{\dnu vs. \numax for all \total stars in our catalog, where the \newdet new detections are shown in blue squares. The solid, orange line shows the best--fit relation between \numax and \dnu. The bottom panel shows the residuals.}
\label{fig:dnu_numax}
\end{center}
\end{figure}

\begin{figure}[t!]
    \centering
    \includegraphics[width=\linewidth]{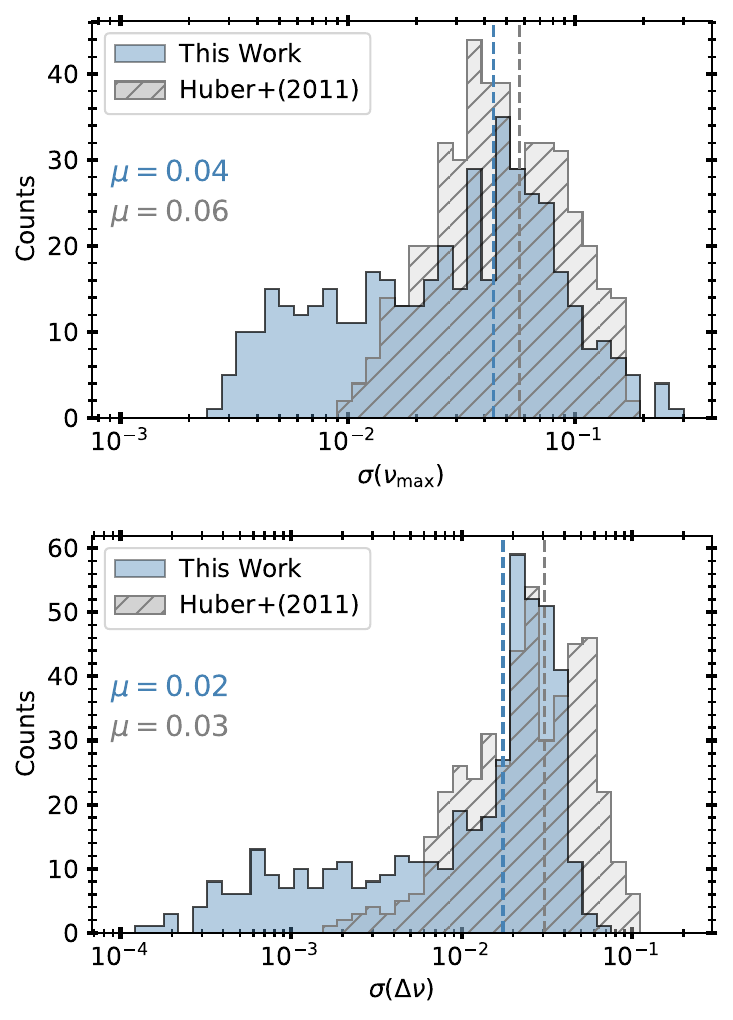}
    \caption{Fractional uncertainty in \numax (top) and \dnu (bottom) for 458 overlapping targets in this work (blue) and H11. The dashed lines and text indicate the mean fractional uncertainty for each sample.}
    \label{fig:frac_err}
\end{figure}

\section{Oscillation Parameters}
\subsection{\dnu -- \numax Relation} \label{sec:dnu-numax}
Our new homogeneous catalog allows us to revisit empirical relations from global oscillation parameters. Figure \ref{fig:dnu_numax} shows the \dnu and \numax relation for our sample. The orange solid line corresponds to the power law relation between \numax and \dnu established for main--sequence and red giant stars \citep{stello_2009, hekker_2009, mosser_2010, hekker_2011b, hekker_2011a}: 
\begin{equation}
    \Delta \nu = \alpha (\nu_{\text{max}}/\mu \text{Hz} )^\beta
\end{equation}
where $\alpha$ = 0.22 and $\beta$ = 0.797 for stars with \numax greater than 300 \muhz (H11). We re--derive the relation and obtain similar fitting parameters: $\alpha=0.216\pm0.005$ and $\beta=0.801\pm0.003$, consistent with previous results. The mean offset is $0.69\pm0.16\%$ with a scatter of 4.4\%.

As seen in the bottom panel of Figure \ref{fig:dnu_numax}, the scatter increases slightly for stars with \numax $\sim600-1000$ \muhz. This parameter space consists of stars where the pressure modes ($p-$modes) and gravity modes ($g-$modes) overlap and create ``mixed modes" \citep{Dziembowski_2001}. The coupling of these two modes causes mixed modes to be shifted from their original frequency spacing, yielding multiple frequencies per radial order \citep{Aizenman_1977}. Mixed modes only occur for $l>0$ modes; since \pysyd calculates the ACF over all modes (instead of just $l=0$ modes), the estimation of \dnu\ becomes more uncertain due to the coupling of $p-$modes and $g-$modes. 

The median uncertainty in \numax and \dnu are $3.4\%$ and $1.9\%$, respectively; only 14 stars have an offset greater than $10\%$. The \newdet new detections have median uncertainty in \numax and \dnu of $5.0\%$ and $2.7\%$, respectively. Figure \ref{fig:frac_err} compares the fractional error in \numax and \dnu in this work with the short--cadence targets in H11. Of the 542 detections with short--cadence data in H11, 463 overlap with our sample; of the remaining 79 targets in H11 not found in our catalog, 72 are confirmed detections with long--cadence data in \cite{Yu_2018} and are therefore not found in our sample by construction, and seven are non--detections. The top and bottom panels in Figure \ref{fig:frac_err} show the fractional error in \numax and \dnu, respectively. The mean error in \numax between our work and H11 is 4\% and 6\%, respectively, while for \dnu is 2\% and 3\%. Our measurement uncertainties in \numax and \dnu peak at lower values compared to those from H11. The stars with error in \dnu less than 0.5\% have narrow peaks in the ACF that produce a small error in \dnu; this is caused by the availability of more data: H11 only used data from Q$0-4$ while our sample contains data from all available quarters. Similarly, the 44 stars with less than 0.5\% uncertainty in \numax have high resolution data (ie. the modes are clearly distinguishable in the \'echelle diagram) where all except for three stars had data available after Q4.

\begin{figure}[t!]
\begin{center}
\includegraphics[width=\linewidth,angle=0]{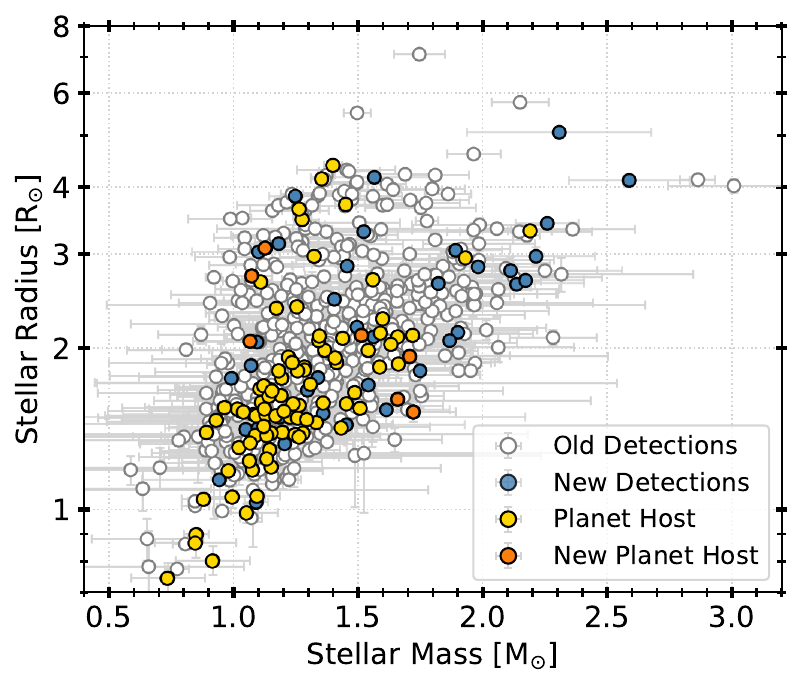}
\caption{Derived stellar masses and radii using Equations \ref{eq:sr_radius} and \ref{eq:sr_mass} for our sample, including new detections in blue, planet hosts in yellow, and new planet hosts in orange. 70 out of \total stars are not shown given no measurement of \numax, \dnu, or \teff.}
\label{fig:mass_radius}
\end{center}
\end{figure}

\subsection{Fundamental Parameters}

The global oscillation parameters \numax\ and \dnu\ can be combined with effective temperature to derive stellar radius and mass using the following scaling relations \citep{ulrich_1986, brown_1991,kjeldsen_bedding},

\begin{equation}
    \Bigg(\frac{\mathrm{R}}{\mathrm{R}_\odot}\Bigg) \simeq \Bigg( \frac{\nu_{\mathrm{max}}}{\nu_{\mathrm{max \odot}}} \Bigg) 
    \Bigg(\frac{\Delta \nu}{\Delta \nu_{\odot}} \Bigg)^{-2}
    \Bigg(\frac{\mathrm{T}_\mathrm{eff}}{\mathrm{T}_{\mathrm{eff}\odot}} \Bigg)^{0.5}
    \label{eq:sr_radius}
\end{equation}

\begin{equation}
    \Bigg(\frac{\mathrm{M}}{\mathrm{M}_\odot}\Bigg) \simeq \Bigg( \frac{\nu_{\mathrm{max}}}{\nu_{\mathrm{max \odot}}} \Bigg) ^3
    \Bigg(\frac{\Delta \nu}{\Delta \nu_{\odot}} \Bigg)^{-4}
    \Bigg(\frac{\mathrm{T}_\mathrm{eff}}{\mathrm{T}_{\mathrm{eff}\odot}} \Bigg)^{1.5}
    \label{eq:sr_mass}
\end{equation}

where $\nu_{\mathrm{max \odot}} = 3090\pm \; 30 \; {\rm \mu Hz}$, $\Delta \nu_{\odot} = 135.1 \pm 0.1 \; {\rm \mu Hz}$, and $\mathrm{T}_{\mathrm{eff}\odot} = 5777 \; {\rm K}$. Values for effective temperature were obtained from \citet{berger_2020}. In Figure \ref{fig:mass_radius}, we show the seismically derived stellar radii and masses for our sample, where 70 stars are not shown given that they do not have either \numax, \dnu, or \teff values. The median error in radius and mass is 2.7\% and 10.5\%, respectively, with radii between $0.7-7.1 \; {\rm R}_\odot$ and masses between $0.6-3.0 \; {\rm M}_\odot$. In Figure \ref{fig:mass_radius}, nine stars have $M < 0.8 \; {\rm M}_\odot$, of which all except one have mass uncertainty above 20\%. There are 18 stars with $M> 2.1 \; {\rm M}_\odot$, however all except four have mass uncertainties above 5\%, and 15 out of 18 have only 30 days of time--series data which could be responsible for the low data quality and uncertain asteroseismic measurements.

Asteroseismic masses and radii are sensitive to the accuracy of the effective temperatures. Since spectroscopically derived temperatures from \texttt{Specmatch} are available for a limited number of stars, we chose to use isochrone based effective temperatures from \cite{berger_2020} in Equations \ref{eq:sr_radius} and \ref{eq:sr_mass} to ensure homogeneity. \cite{berger_2020} validated their derived \teff against interferometric and spectroscopic temperatures (see Sections 3.1 and 4.5, respectively). They found an offset of $-3\%$ to $1\%$ when comparing to spectroscopic temperatures compiled in \cite{Mathur_2017}, and within $2\%$ when compared to interferometric temperatures from \cite{Boyajian2013} and \cite{Huang2015}.

\begin{figure}[t!]
\begin{center}
\includegraphics[width=0.9\linewidth,angle=0]{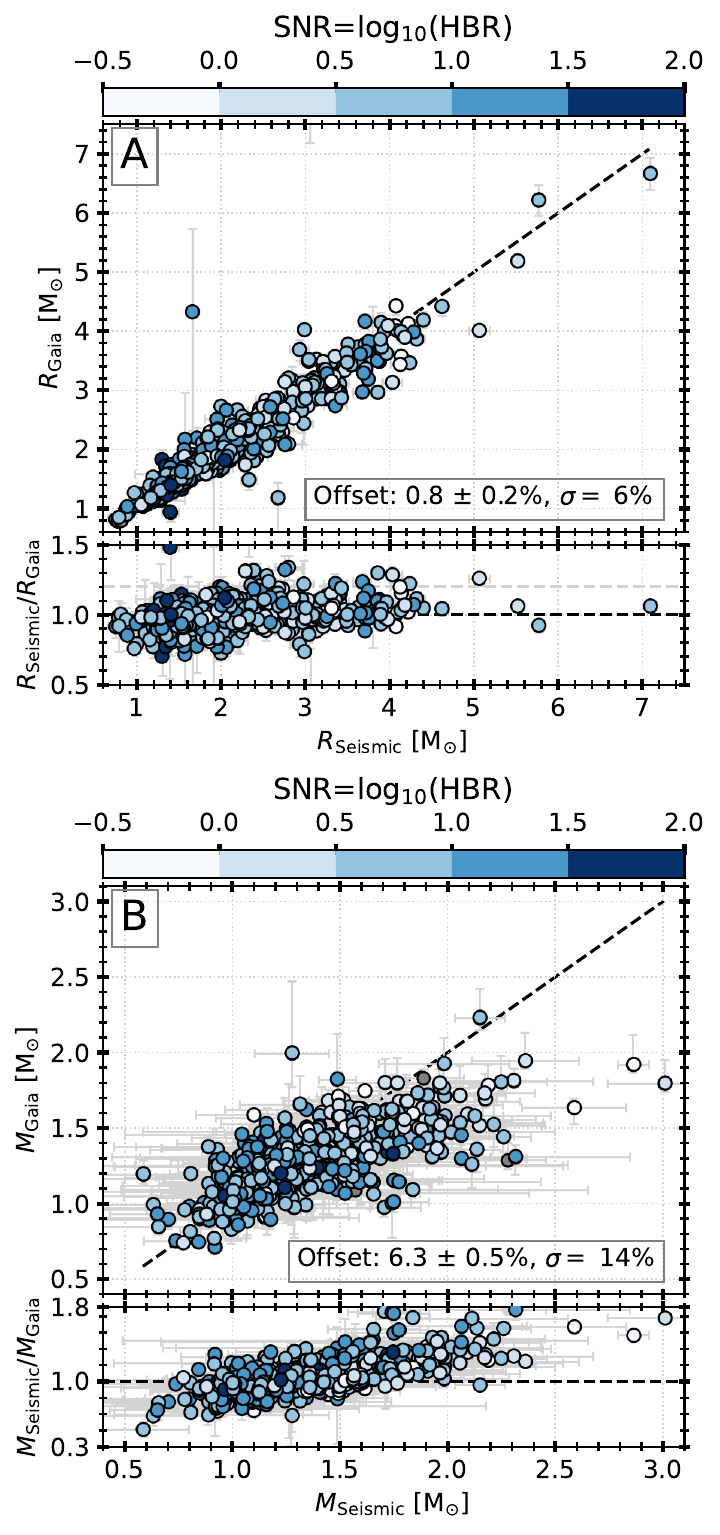}
\caption{Comparison of radii (\textit{top}) and masses (\textit{bottom}) derived with asteroseismic scaling relations (x--axis) and with \gaia measurements (y--axis). Residuals (ie. ratio between the seismic and \gaia derived quantities) are shown in bottom panels. The color--coding denotes the height--to--background ratio, defined as the ratio between the amplitudes of oscillation and background. The text indicates the mean offset and error between the two measurements and the scatter, $\sigma$.} \label{fig:compare_radius}
\end{center}
\end{figure}

In Figure \ref{fig:compare_radius}, we compare our masses and radii with those derived from isochrone fitting using \gaia parallaxes from \cite{berger_2020}. The colour map represents the height--to--background ratio (HBR) of the star at \numax where the height is the smoothed amplitude at \numax \citep{Mosser2012}, and the background is fit with a Harvey model \citep{Harvey_1985}. Therefore, HBR is the ratio between the height and background at \numax with white noise removed from the background fit,

\begin{equation}
    \textrm{HBR} = \frac{H_{\nu_{max}}}{B_{\nu_{max}}}
    \label{eq:snr}
\end{equation}

The mean offset and error for the radius comparison (Panel A in Figure \ref{fig:compare_radius}) is $0.8 \pm 0.2$\%  and scatter of 6\%. \cite{zinn_2019} performed a similar analysis comparing the \gaia radii to seismic radii for main--sequence and subgiant stars, and achieved a residual median and scatter of \app1\% and \app4\%, respectively. Although we achieve similar precision, our sample mostly contains low SNR stars, where 66\% of the sample has HBR below 10. To investigate the source of outliers, we inspect the re--normalized unit weight--error (RUWE) from \gaia DR3. RUWE is the magnitude and colour--independent re--normalization of the astrometric $\chi^2$ is sensitive to close binaries \citep[e.g.,][]{evans_2018, gaia_2018, Lindegren_2018, berger_2020, gaia_dr3}. Stars with RUWE $\gtrsim$ 1.5 are more likely to have a companion. Of the outliers, the star at ($R_{\mathrm{S}}, R_{\mathrm{G}}) \simeq (1.6, 4.3)$ is KIC ID 5513648 with a RUWE of 14 in \gaia DR2 (a \gaia DR3 value is not available for this star), the star at ($R_{\mathrm{S}}, R_{\mathrm{G}}) \sim (5, 4)$ is $\mathrm{KIC \; ID \; 6627507}$, a subgiant with only one month of data, and the star at ($R_{\mathrm{S}}, R_{\mathrm{G}}) \sim (2.6, 1.2)$ is KIC ID 6863041 with a RUWE of 4.9. There is one star outside of the plot limits, KIC ID 11558593 at ($R_{\mathrm{S}}, R_{\mathrm{G}}) \sim (3.0, 8.5)$ which has a RUWE of 7.8. Three of the four stars with greater than 50\% difference between the two radii have RUWE $> 4.9$ (the fourth star is KIC ID 5513648 and does not have a RUWE measurement in \gaia DR3); this suggests that the \gaia radii are most likely untrustworthy since the photometry will be affected by companions. Targets with an offset between 20--50\% show no correlation with \numax, \kp, RUWE, effective temperature, and \logg. We conclude that these differences are due to the difficulty of measuring accurate radii from isochrone fitting alone.

Figure \ref{fig:compare_radius}B compares seismic mass and masses from isochrone fitting, or \gaia masses. The mean offset is $6.3 \pm 0.5$\% with a scatter of 14\%, where 22\% of the targets have greater than 20\% offset between the seismic and \gaia masses; however of these, 88\% have RUWE below 1.5 suggesting that contamination in the photometry is not responsible for inaccurate mass measurements. Furthermore, there is an overall trend where seismic masses are systematically higher than \gaia masses for our mass range as seen in the residuals. However, there is no correlation between this large offset in mass and stellar parameters (eg., \teff, radii, RUWE, HBR, \kp, nor error in \numax, \dnu, \teff, or stellar mass) suggesting that masses from isochrone fitting are uncertain.

\begin{figure}[t!]
\begin{center}
\includegraphics[width=0.9\linewidth,angle=0]{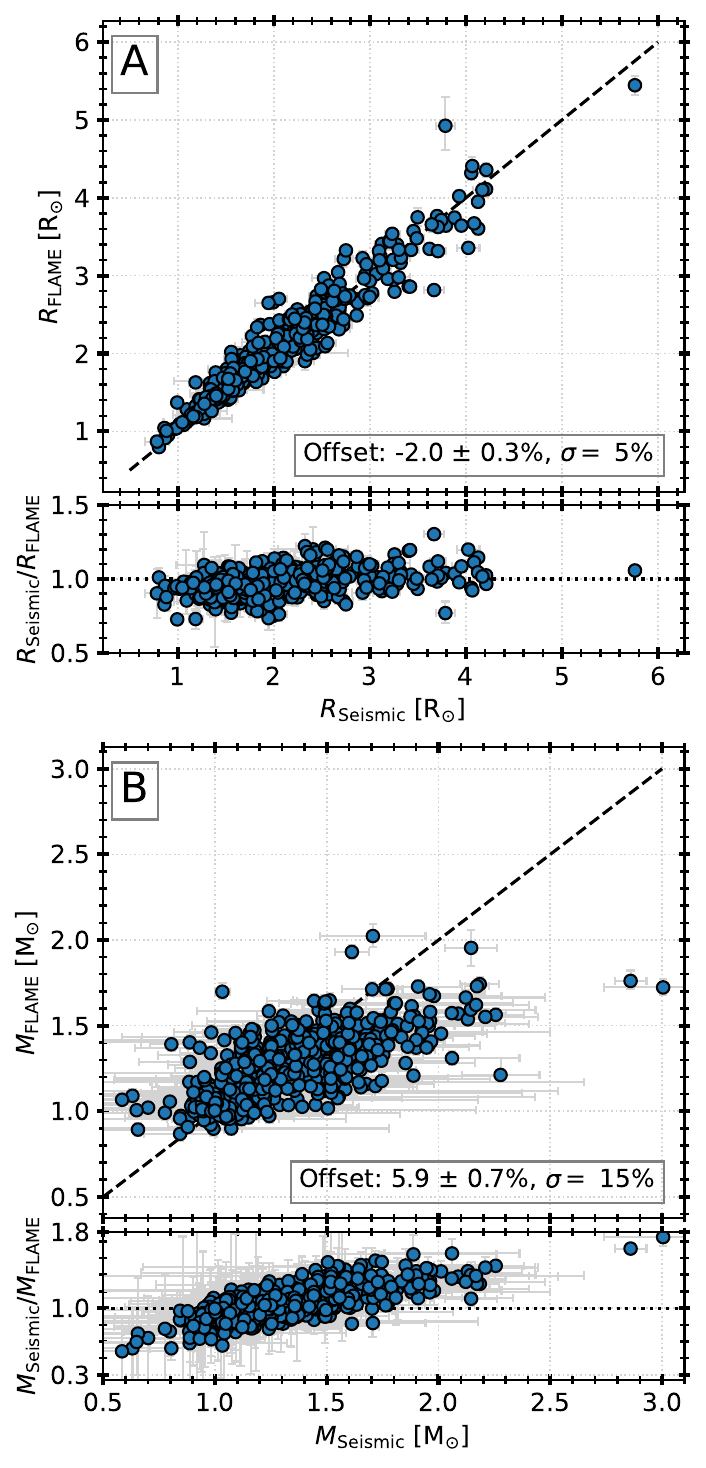}
\caption{Comparison of radii (\textit{top}) and masses (\textit{bottom}) derived with asteroseismic scaling relations (x--axis) and \gaia DR3 FLAME measurements (y--axis). Residuals are shown in the bottom panels. The text indicates the mean offset and error between the two measurements and the scatter, $\sigma$.} \label{fig:flame}
\end{center}
\end{figure}

Figure \ref{fig:flame} compares our stellar radii and masses to those estimated by the Gaia Final Luminosity Age Mass Estimator (FLAME) pipeline \citep{Hidalgo2018, Creevey2023}. We used \texttt{TOPCAT} to match our sample to those from \gaia DR3 with a finding radius of $2''$. We also ensure good quality flags in the FLAME radii and masses (ie. first digit in \texttt{Flags-Flame} is 0). For our sample of \total targets, FLAME radii and masses are available for 584 and 581 targets, respectively, as compared to 695 targets with both seismic measurements and isochrone derived radii and masses from \cite{berger_2020}. For both radius and mass, we observe distributions similar to those in Figure \ref{fig:compare_radius}, with a mean offset of $-2.0 \pm 0.3\%$ with a scatter of 5\% in radius, and a mean offset of $5.9\pm0.7\%$ with a scatter of 15\% in mass.



\subsection{Oscillation Amplitudes}

We calculated oscillation amplitudes per radial mode using the following relation from \cite{Kjeldsen_2008},
\begin{equation}
    A = \sqrt{\frac{\text{PSD} \cdot \Delta \nu }{c}}
    \label{eq:amp}
\end{equation}
where PSD is the maximum power spectral density of the smoothed oscillation envelope in ppm$^2$/\muhz returned by \pysyd at \numax, \dnu is the frequency separation of the star in \muhz, and $c$ is the normalization factor of 3.04 \citep{Kjeldsen_2008, Huber_2011}. 

\begin{figure}[t!]
\begin{center}
\includegraphics[width=\linewidth,angle=0]{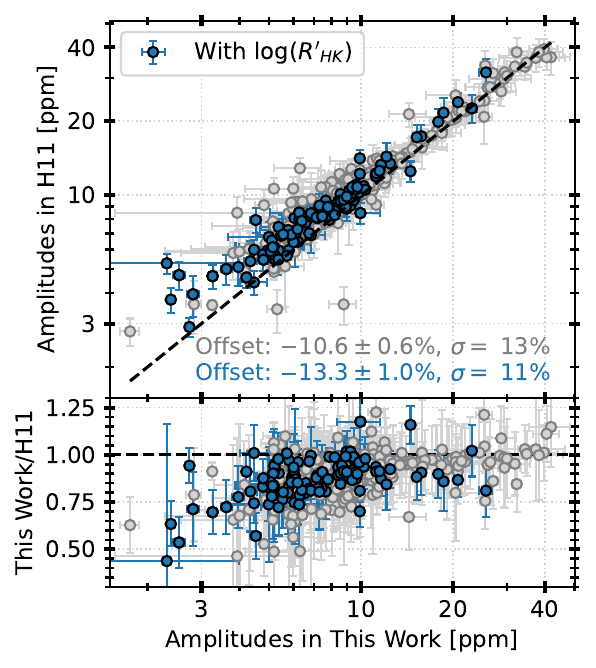}
\caption{Comparison of amplitudes derived in this work to those from \cite{Huber_2011} for 447 targets in grey. The blue points show the 132 targets for which \rhk values are available in \hires. The text indicates the mean offset with error, and scatter in both distributions. The dashed line indicates unity. The bottom panel shows the residual.}
\label{fig:compare_amps}
\end{center}
\end{figure}

Figure \ref{fig:compare_amps} compares our amplitude measurements to those from H11. We see a clear offset in the amplitudes from this work and H11, with mean offsets of $10.6\pm0.6\%$ and $13.3\pm1.0\%$, respectively. We suspect that these offsets are a result of improved SNR due to the longer timeseries used in this paper, and differences between the original pipeline version used by H11 and \pysyd. Specifically, \pysyd fits for the white noise level in the background modeling step, rather than using a fixed white noise value (ie. average amplitude at high frequencies) as was used in H11. In addition, \pysyd optimizes the specific Harvey model used where the options are `regular', `second', or `fourth'; the regular model includes the second and fourth order terms, the second model includes only the second order term, and the fourth model includes only the fourth order term. The pipeline in H11 only used the regular model with both the second and fourth order terms, and this step was therefore not optimized for each star.

\begin{figure*}[t!]
\begin{center}
\includegraphics[width=\linewidth,angle=0]{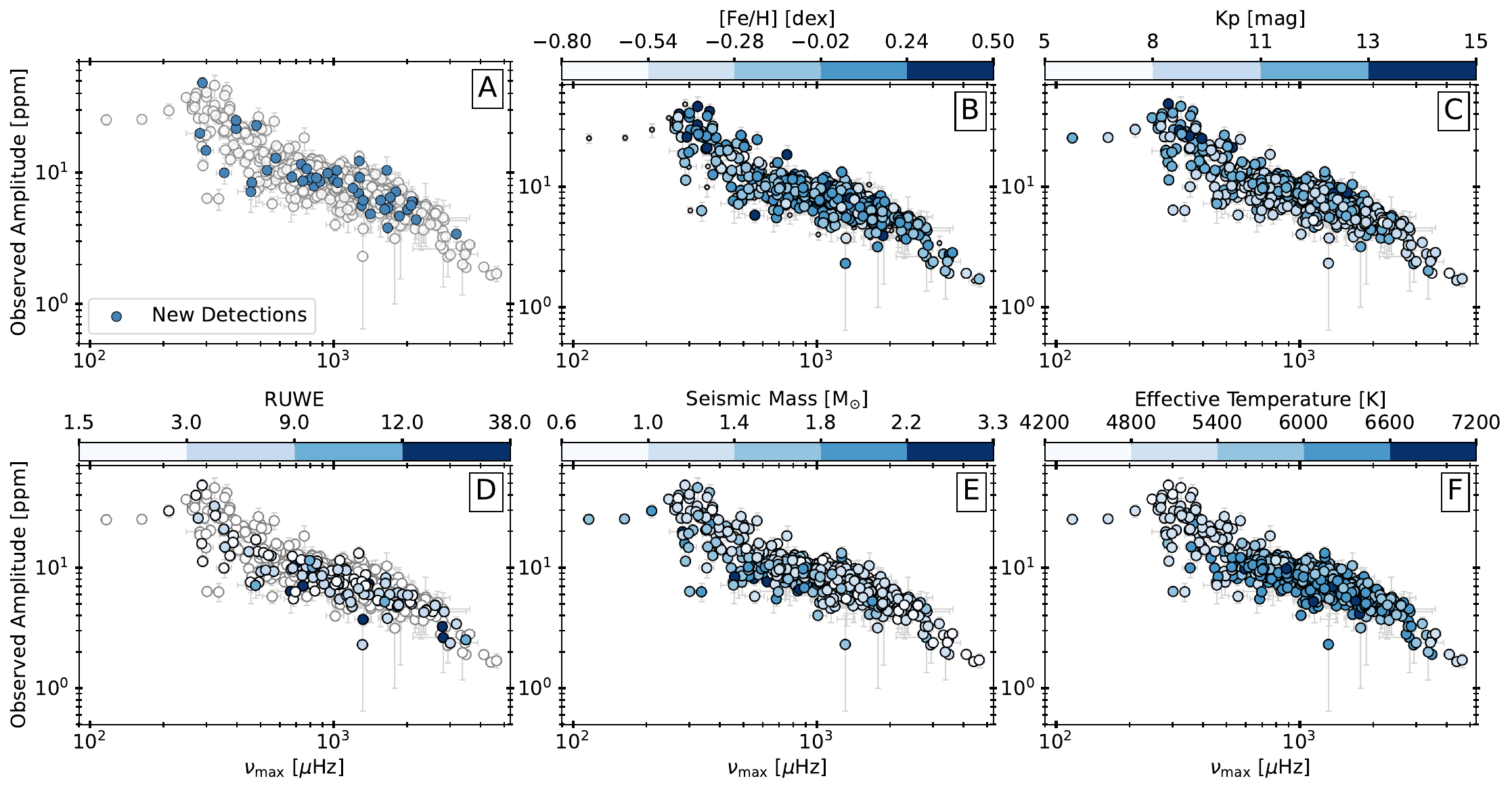}
\caption{Observed amplitude at \numax as a function of frequency of maximum oscillation. Panel A shows the new detections in blue against the overall sample in grey. Panels B--F are coloured by stellar properties: metallicity (B), \kp (C), \gaia DR3 $\mathrm{RUWE \geq 1.5}$ (D), seismic mass (E), and \gaia effective temperature from \cite{berger_2020} (F).}
\label{fig:all_amp_vs_numax}
\end{center}
\end{figure*}

In Figure \ref{fig:all_amp_vs_numax}A, we plot the observed amplitude as a function of measured \numax. Note that 47 stars are missing from this plot given no measurement of \numax or \dnu. 
The star at (\numax, $A_\mathrm{obs}$ ) $\sim$ (100, 17) -- KIC ID 6470149 -- has a clear oscillation pattern, but its low resolution makes the \numax measurement difficult and could be responsible for its deviation from the distribution. Similarly, KIC ID 9390670 at (\numax, $A_\mathrm{obs}$ ) $\sim$ (1304, 2.3) is a low SNR detection with an uncertain \numax, and a greater than 50\% error in observed amplitude. The observed change in slope at \app$1500$ \muhz is consistent with previous studies (e.g., H13). 

Similar to previous studies (e.g., H11), the amplitude--\numax\ relation shows significantly more scatter than expected from measurement errors. To investigate possible correlations with other parameters, Figure \ref{fig:all_amp_vs_numax}B shows the same distribution coloured by the metallicity for each star, where spectroscopic metallicities are available for 73\% of our sample from three sources: \cite{Bruntt_2012, Buchhave_2015, Serenelli_2017}. We find no gradient or pattern for main--sequence and subgiant stars in this parameter space. This is similar to the results from \cite{Yu_2018} for red giant stars, but different than their results for red clump stars, where they found that metal--rich red clump stars oscillate with larger amplitudes than metal--poor stars at a given \numax, \dnu, and \teff (see Figure 12 in \cite{Yu_2018}). 

Figure \ref{fig:all_amp_vs_numax}C shows the distribution colour--coded by \kp, with the lack of correlation confirming that the spread is not due to an inaccurate white noise correction. In Figure \ref{fig:all_amp_vs_numax}D, we show the relationship between RUWE from \gaia DR3 and amplitude at \numax. In our sample, 122 stars (or 17\% of the sample) have RUWE above 1.5 which are coloured in Figure \ref{fig:all_amp_vs_numax}D. A close companion would dilute the amplitude (due to extra light), which could yield an underestimated amplitude, where low amplitude corresponds to higher stellar activity \citep[e.g.,][]{chaplin_2000, komm_2000, chaplin_2011}. As seen in Panel D, stars with RUWE $\geq 1.5$ follow the overall distribution of our sample with no outliers. Figure \ref{fig:all_amp_vs_numax}E shows the relationship between seismic mass and amplitude, where mass is derived using Equation \ref{eq:sr_mass}. We find that lower mass stars have higher amplitude at a given \numax on average while the opposite is true for higher mass stars, consistent with H11 and \cite{Yu_2018}. Lastly, Figure \ref{fig:all_amp_vs_numax}F shows no correlation between effective temperature and amplitude at a fixed \numax.

\subsection{A New Oscillation Amplitude Scaling Relation including Chromospheric Activity} \label{sec:relation}

H11 derived a scaling relation to predict oscillation amplitude given stellar mass, radius, and temperature. We revise their scaling relation by including available chromospheric activity measurements for our sample. To do this, we derive a modified amplitude scaling relation including the \rhk term normalized to solar \rhk value of $-4.94$ \citep{Egeland2017},

\begin{equation}
     \frac{A}{A_\odot} = \frac{L^s (\log(R'_{\rm HK})/\log(R'_{\rm HK})_\odot)^u}{M^t T^{r-1}_\mathrm{eff}c_K(T_\mathrm{eff})}, \quad c_K = \Bigg(\frac{T_\mathrm{eff}}{5934}\Bigg)^{0.8}
     \label{eq:exp_amp}
\end{equation}
where the solar reference amplitude $A_\odot$ is 3.6 ppm, and $c_K$ is the bolometric correction factor as a function of effective temperature \citep{Ballot_2011}. 

We re--derived the solar reference values for \pysyd following H11 by using 112 30--day subsets of data from the VIRGO instrument \citep{frohlich1997}, after processing the data following methods described in Section \ref{sec:data}. We measure a mean oscillation amplitude of $4.57\pm0.33 \rm \; ppm$, $\nu_{\max} = 3080\pm30 \; \mu \rm Hz$, and $\delta \nu = 135.1\pm0.1 \rm \; \mu Hz$. Given VIRGO's green channel ($\lambda = 500 \rm \; nm$), we convert our measured amplitude to a solar bolometric amplitude of $3.66\pm0.27 \rm \; ppm$ using methods outlined in \cite{kjeldsen_bedding}. Our reference values are consistent with H11 and the solar amplitude from \cite{Michel2009}, so we continue to use $A_\odot=3.6 \rm \; ppm$ in the following analysis for consistency.

To predict amplitudes, we use asteroseismic radii and masses derived with both photometric \teff from \cite{berger_2020}, and spectroscopic \teff from \texttt{Specmatch}. Table \ref{tb:rphk} includes spectroscopic parameters (\teff, \logg, [Fe/H]), $B-V$, \rhk, $\sigma({\rm \log} {R'_{\rm HK}})$ and re--derived seismic masses and radii for 127 stars. We estimate uncertainties on \rhk via bootstrapping by perturbing the spectroscopic quantities with typical uncertainties. For the $S-$index, we assume a typical uncertainty of 1.6\% as found by \cite{gomesdasilva_2021}. The mean relative uncertainty in \rhk is therefore 3\%, similar to \cite{gomesdasilva_2021}.

There is an offset of $0.5\pm0.1\%$ between the isochrone derived \teff from \cite{berger_2020} and \texttt{Specmatch} \teff, with a scatter of $1.9\%$. Of the 172 stars with available \rhk values, $T_{\rm eff,iso}$, and $T_{\rm eff,spec}$, 40 stars (or 23\% of the sample) have $\Delta T_{\rm eff}>100 \; {\rm K}$, but 14 of them are hot stars, with $T_{\rm eff,iso}>6400 \; {\rm K}$. On average, the difference between the two \teff measurements is $31 \; \rm{K}$.

\begin{figure*}[t!]
\begin{center}
\includegraphics[width=\linewidth,angle=0]{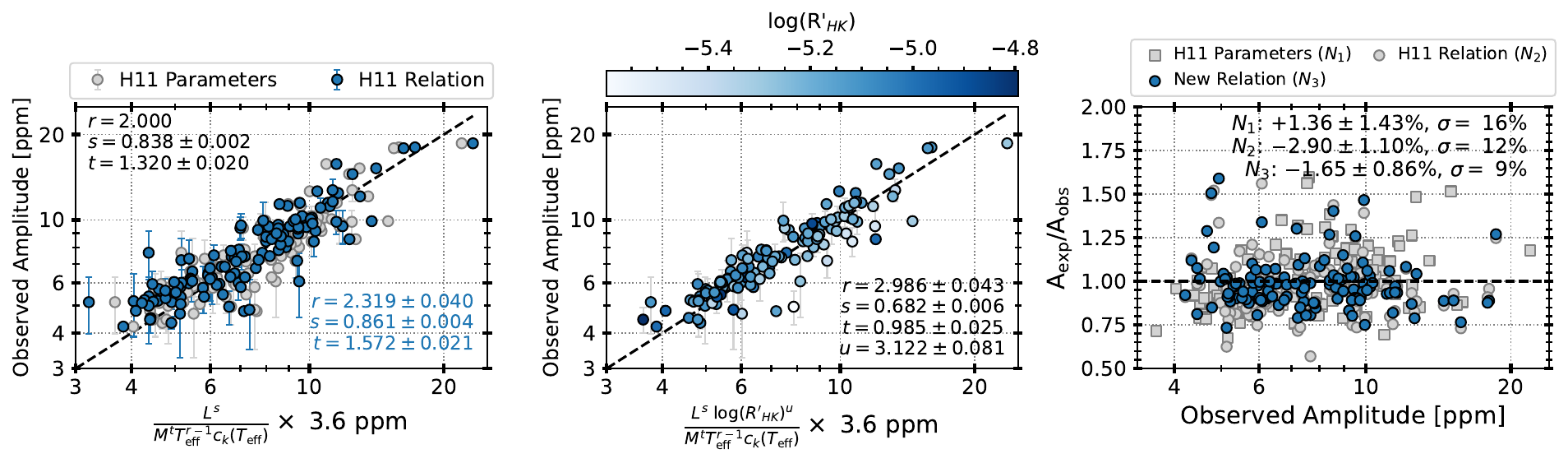}\hfill
\includegraphics[width=\linewidth,angle=0]{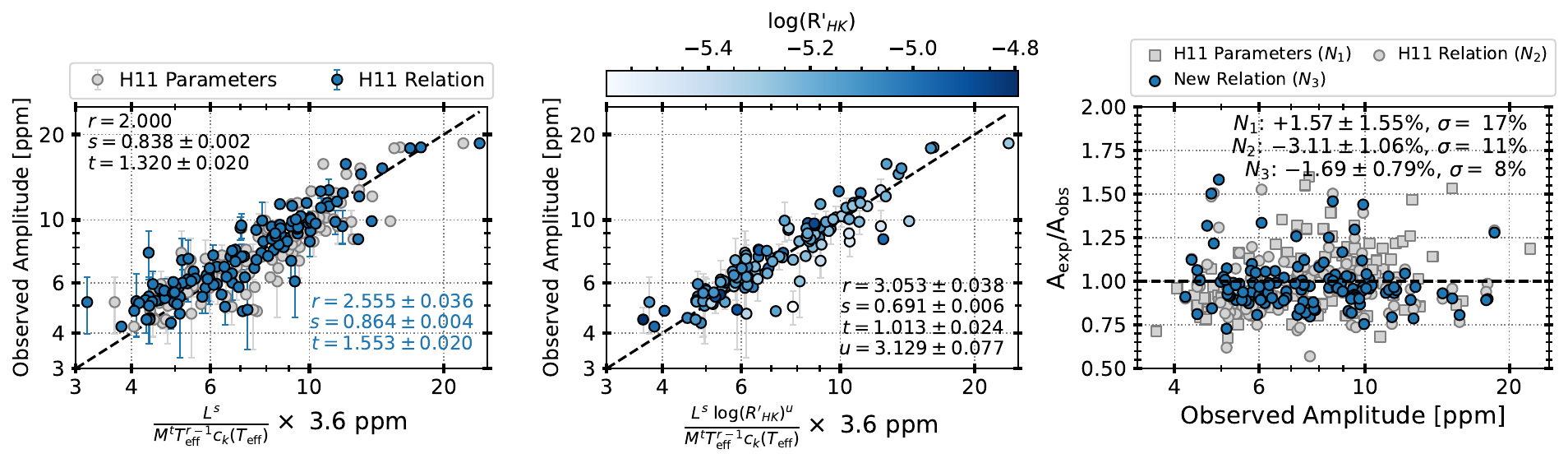}
\caption{Amplitude scaling relations derived with spectroscopic \teff from \texttt{Specmatch} (top panels) and isochrone based \teff from \cite{berger_2020} (bottom panels). \textit{Left:} Observed amplitude as a function of the expected amplitude derived using Equation \ref{eq:exp_amp} without an activity term. The grey points show the expected amplitudes using best fit parameters from H11 and the blue points show the expected amplitudes with the relation from H11 but with new best--fit parameters, $r$, $s$, and $t$. The dashed line shows the one--to--one relation. The text indicates the best--fit parameters for both distributions. 
\textit{Middle:} Same as left panel but including an activity term, coloured by the \rhk index from \hires. The annotated text indicates the best--fit parameters, and the dashed line shows the one--to--one relation.
\textit{Right:} Residuals between observed and predicted amplitudes using the relation and best--fit parameters from H11 (grey circles), the expected amplitudes derived using H11 relation but re--derived best--fit parameters (grey squares), and the new relation from Equation \ref{eq:exp_amp} (the blue circles). The annotated text indicates the mean offset with error and the scatter.}

\label{fig:exp_amp}
\end{center}
\end{figure*}

We limit our sample to targets between $A_{\rm obs}=[4, 20] \; \rm{ppm}$, and use both spectroscopic \teff from \texttt{Specmatch} ($T_{\rm eff, \; spec}$) and isochrone based \teff from \cite{berger_2020} ($T_{\rm eff, \; iso}$) in Equation \ref{eq:exp_amp} to calculate expected amplitudes. In Figure \ref{fig:exp_amp}, we show the results from spectroscopic \teff in the top panel and isochrone based \teff in the bottom panel for 127 stars. In the left panel of Figure \ref{fig:exp_amp}, we derive the expected amplitudes with our observed amplitudes and H11 scaling relation, similar to Equation \ref{eq:exp_amp}, but without the \rhk term (e.g., $u=0$). We compare the best--fit parameters from H11 (grey points), with new best--fit parameters derived using an optimization routine. The new best--fit parameters with $T_{\rm eff, \; spec}$ and $T_{\rm eff, \; iso}$ are shown in Figure \ref{fig:exp_amp} left panel and listed in the first column in Table \ref{tb:tb-fit-values}; both sets of best--fit parameters are similar to those in H11: $r=2$, $s = 0.838 \; \pm \; 0.002$, and $t = 1.32 \; \pm \; 0.02$.

In the middle panel of Figure \ref{fig:exp_amp}, we derive expected amplitudes for main--sequence and subgiant stars (\teff $ \in [4900,6700]$ K, \logg $\in [3.2, 4.5]$ dex, and \rhk $\in [-5.6, -4.8]$) with four fit parameters, $r$, $s$, $t$, and $u$. The best--fit parameters are listed in the right column of Table \ref{tb:tb-fit-values}. There are eleven stars\footnote{KIC IDs 3123191, 3967859, 5856836, 7510397, 7510397, 8018599, 8420801, 8561221, 11389437, 11389437, 11453915} with greater than 25\% offset between the expected and observed amplitudes using both $T_{\rm eff, \; spec}$ and $T_{\rm eff, \; iso}$. Two stars (KIC ID 7510397 and 3123191) have RUWE above 1.5, while the remaining stars have RUWE below 1.1. These targets show no correlation between their seismic and stellar parameters (ie. \numax, \dnu, \teff, \kp, \prot, [Fe/H], HBR) and their large offset.

\begin{deluxetable}{c|c|c}[t!]
    \tablewidth{0pt}
    \tablecaption{Best--fit parameters in Equation \ref{eq:exp_amp} using the old scaling relation from H11 without the activity term (ie. $u=0$ in Equation \ref{eq:exp_amp}) and the new scaling relation with the activity term. The mean offset with error, scatter, and BIC (Equation \ref{eq:bic}) is included for both models for reference. The best--fit parameters are listed using both spectroscopic and isochrone based effective temperatures, $T_{\rm eff, \; spec}$ and $T_{\rm eff, \; iso}$.  \label{tb:tb-fit-values}}
    \tablehead{ & \multicolumn{2}{c}{Model}  \\\hline
    \multicolumn{1}{c|}{Parameter} & \multicolumn{1}{c|}{H11 Relation} & \multicolumn{1}{c}{New Relation} }
    \startdata
    \hline \hline
    \multicolumn{3}{c}{Best Fit Parameters with $T_{\rm eff,spec}$} \\\hline\hline
    $r$ & $2.319\pm0.040$ & $2.986\pm0.043$ \\
    $s$ & $0.861\pm0.004$ & $0.682\pm0.006$ \\
    $t$ & $1.572\pm0.021$ & $0.985\pm0.025$\\
    $u$ &  0 & $3.122\pm0.081$ \\\hline
    Offset & $-2.90\pm1.10$& $-1.65\pm0.86$ \\
    Scatter & $12\%$& $9\%$ \\\hline
    BIC & 89.2  & 73.3 \\\hline\hline
    \multicolumn{3}{c}{Best Fit Parameters with $T_{\rm eff,iso}$} \\\hline\hline
    $r$ & $2.555\pm0.036$ & $3.053\pm0.038$ \\
    $s$ & $0.864\pm0.004$ & $0.691\pm0.006$ \\
    $t$ & $1.553\pm0.020$ & $1.013\pm0.024$\\
    $u$ &  0 & $3.129\pm0.077$ \\\hline
    Offset & $-3.11\pm1.06$& $-1.69\pm0.79$ \\
    Scatter & $11\%$& $8\%$ \\\hline
    BIC & 86.1 & 68.4 \\
    \enddata
\end{deluxetable}

In the right panel of Figure \ref{fig:exp_amp}, we compare the new scaling relation in this work with the relation and best--fit parameters from H11, and relation from H11 but re--derived best--fit parameters. Using the new relation, we achieve an average offset of $-1.65\pm0.86$\% and scatter of $\sigma=9\%$ with $T_{\rm eff, \; spec}$, and $-1.69\pm0.79$\% and $\sigma=8\%$ with $T_{\rm eff, \; iso}$. Using H11 relation and best--fit parameters, we obtain an offset of $-2.90\pm1.10$\% and scatter of $\sigma=12\%$ with $T_{\rm eff, \; spec}$, and $-3.11\pm1.06$\% and $\sigma=11\%$ with $T_{\rm eff, \; iso}$. A possible source of scatter could be caused by the metallicity and temperature dependence on $c$ in Equation \ref{eq:amp} \citep{Ballot_2011}, as well as ignoring the metallicity dependence on the bolometric correction \citep{lund2019}. 


\begin{figure}[t!]
    \centering
    \includegraphics[width=\linewidth]{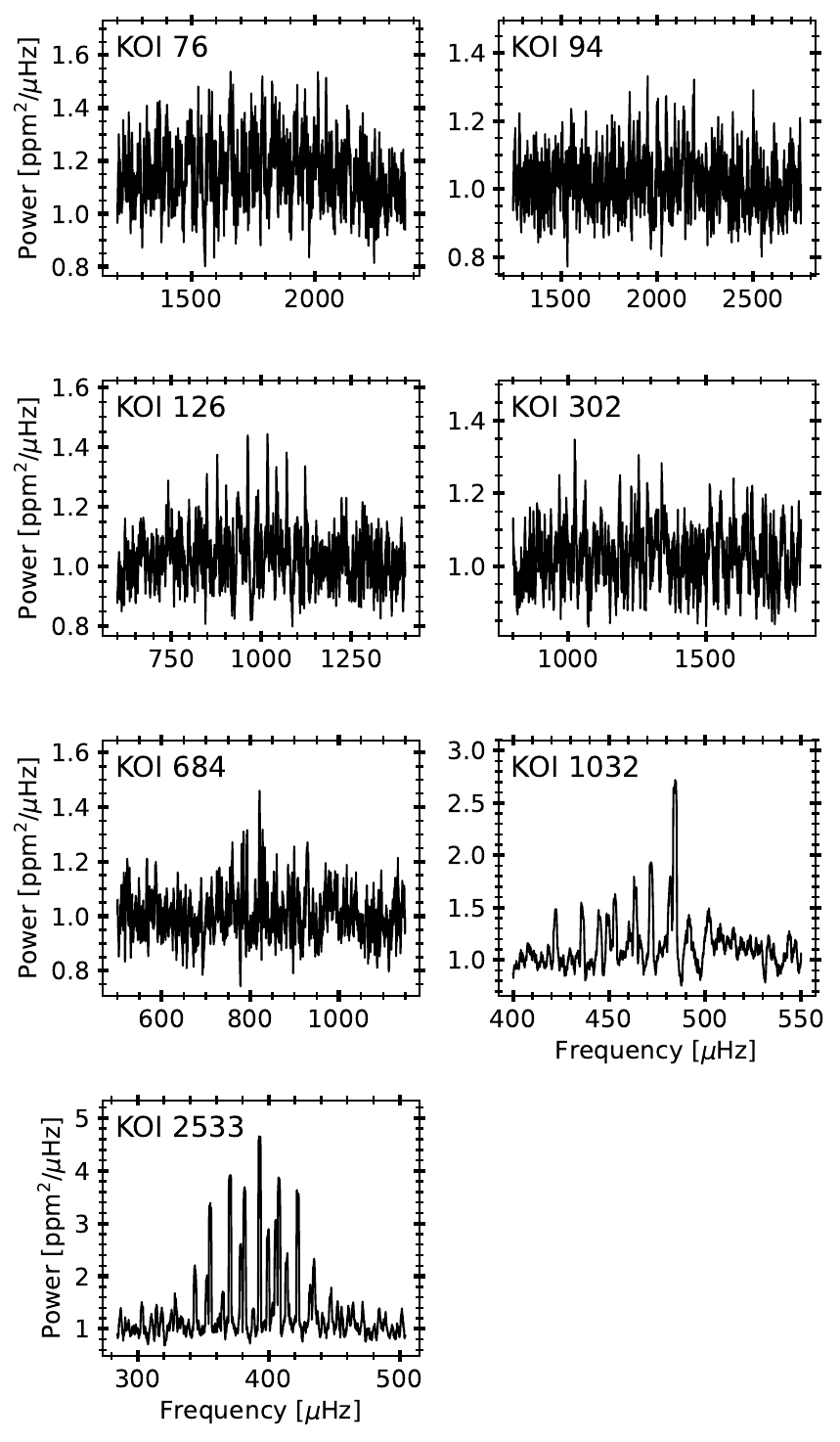}
    \caption{Background divided power spectra for seven \kepler hosts with newly discovered oscillations centered on \numax.}
    \label{fig:spectra_hosts}
\end{figure}

\begin{figure*}[t!]
\begin{center}
\includegraphics[width=\linewidth,angle=0]{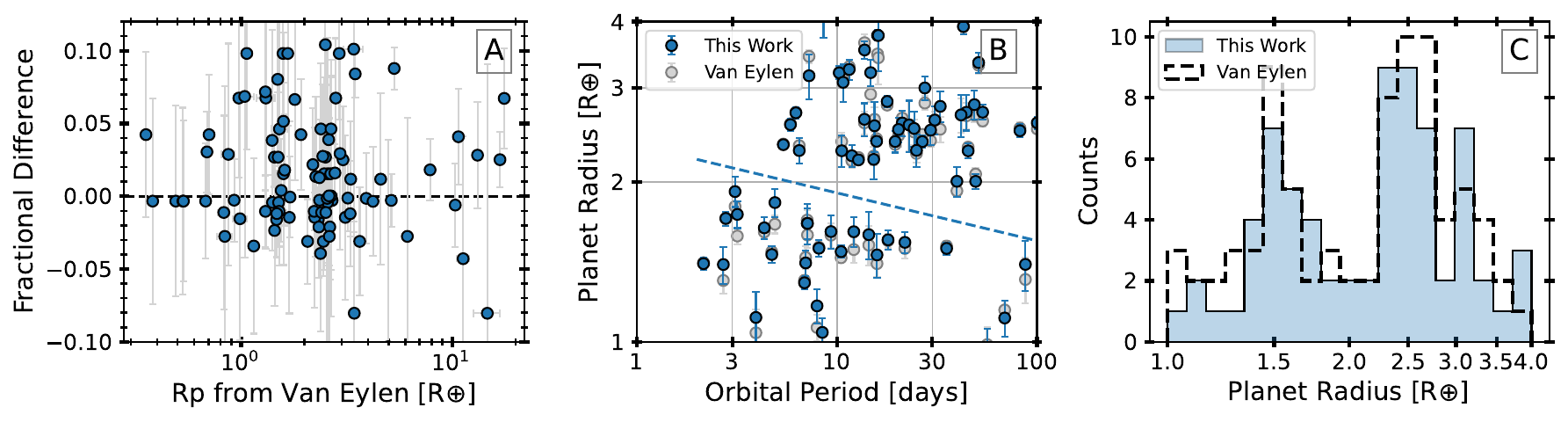}
\caption{\textit{Panel A:} Fractional difference between planet radii from asteroseismology and those from \cite{vaneylen_2018} (VE18): $(R_{\rm p, \; this \; work} - R_{\rm p, \; VE18})/R_{\rm p, \; VE18}$. \textit{Panel B:} planet radii on period--radius plane, where the grey and blue points are planet radii from VE18 and asteroseismology, respectively, and dashed line indicates the radius valley, with best--fit parameters ($m=-0.09$ and $\alpha=0.37$) and axes limits from VE18. \textit{Panel C:} distribution of planet radii from asteroseismology (blue), and from VE18 (dashed black).}
\label{fig:planet_rad}
\end{center}
\end{figure*}

We calculated the Bayesian Information Criterion (BIC) to compare amplitude scaling relation models:
\begin{equation}
    \textrm{BIC} = n\ln\Bigg(\frac{\rm{RSS}}{n}\Bigg) + k \ln (n)
    \label{eq:bic}
\end{equation}
where the likelihood function is the mean squared error (the residual sum of squares (RSS) divided by number of data points), $n$ is the number of data points, and $k$ is the number of free parameters \citep[][]{priestley_spectral_1981, hastie01statisticallearning}. For the new relation, $k=4$, but $k=3$ for the old scaling relation. We calculate $\delta_{\rm BIC} = 15.9$ for $T_{\rm eff, \; spec}$ and $\delta_{\rm BIC} = 17.7$ with $T_{\rm eff, \; iso}$, providing significant evidence in favor of the new model \citep{Raftery1995}.

We conclude that the new scaling relation with the activity term is more successful at recovering observed oscillation amplitudes with a lower offset and scatter. The modest improvement compared to the old scaling relation is likely due to measurement uncertainties in the amplitudes, which are now comparable to the residual scatter ($\approx$\,10\%). Our results of an improved oscillation amplitude with an addition of an activity term is similar to results found by \cite{Tiago2014}; they tested two models to predict oscillation amplitude and found that the model with a magnetic activity proxy performed better the model without. Similarly, \cite{Bonanno2019} proposed a new amplitude--luminosity relation with spot fraction as a proxy for magnetic activity using data from subgiant EK Eri; however, while their new expected amplitude improved, it was still under--predicted by a factor of six, and calibrated on one target only.



\section{Discussion}

\subsection{Exoplanet Host Stars}

We searched for solar--like oscillations in all \kepler stars with short--cadence data including those that were given an exoplanet disposition of confirmed or candidate on the NASA Exoplanet Archive. Given the selection criteria, we also included any exoplanet hosts that were also classified as binaries in \cite{kirk_2016}; 65 targets satisfied this criterion. We confirmed solar--like oscillations in 94 of the 110 previously known detections (69 out of 71 in H13, 25 out of 38 in L16, and 1 in \cite{Chontos_2019}).


We also discovered solar--like oscillations in seven new KOIs: KIC ID 2162635 (KOI 1032), 3662838 (KOI 302, Kepler--516), 5897826 (KOI 126, Kepler--469), 6462863 (KOI 94, Kepler--89), 7730747 (KOI 684), 9955262 (KOI 76), and 11074835 (KOI 2533, Kepler--1270). The background corrected power spectrum for the seven new detections is shown in Figure \ref{fig:spectra_hosts}. The power spectra for the 15 planet hosts which we did not confirm as detections are shown in Figure \ref{fig:unconfirmed_dets}; two targets are from H13 and 13 targets are from L16.

Using our revised stellar radii, we re--derived planet radii as follows,
\begin{equation}
    R_\mathrm{p} = \frac{R_\mathrm{p,V}}{R_\mathrm{s,V}}\times R_\mathrm{s,seismic}
    \label{eq:planet_rad}
\end{equation}
$R_\mathrm{p,V}$ and $R_\mathrm{s,V}$ are planet radius and stellar radius from \cite{vaneylen_2018} (hereon VE18), and $R_\mathrm{s,seismic}$ is seismically derived stellar radii in this work. VE18 contains 75 stars hosting 117 planets. Our catalog contains 108 planets around 67 host stars; of the remaining 9 planets not found in our catalog, 7 are around hosts not confirmed as detections in this work, one target does not have short--cadence data available (KIC ID 9642292, Kepler--1392), and one planet is now labeled as a false positive (KOI--5c). Furthermore, while 64 hosts from VE18 are also found in our catalog, five do not have stellar radius measurement available, due to no measurement in \numax, \dnu, or \teff. Therefore, we compare planet radii for 98 planets around 62 hosts to those from VE18.

Figure \ref{fig:planet_rad}A shows the ratio between the planet radii derived using our stellar radii and planet radii from VE18. The mean offset is $1.4 \pm 0.4\%$ with a scatter of 4\%, suggesting good agreement between the two samples. The mean error in planet radius as found by VE18 is 3.9\%, while the mean error in planet radius derived using Equation \ref{eq:planet_rad} is 6.1\%. 

Figure \ref{fig:planet_rad}B shows the planet radii on a period--radius plane, with orbital periods obtained from VE18, and Figure \ref{fig:planet_rad}C shows the distribution of seismically derived planet radii and those from VE18. We observe a gap in the radius distribution at $R\approx2$ \rearth, consistent with the radius valley, a feature in the demographics of short--period ($P<100$ days) planets with radii between $1.3-2.6$ \rearth \citep{owen_wu_2013, fulton17}. Our distribution is similar to VE18 who investigated the radius valley using accurate stellar parameters from asteroseismology, and also consistent with previous studies \citep[e.g.,][]{owen_wu_2013, jin_2014, lopez_fortney_2014, chen_rogers_2016, owen_wu_2017, lopez_rice_2018, vaneylen_2018, berger_2020b}. Leading theories to explain the radius valley include photoevaporation \citep{owen_wu_2017}, core--powered mass--loss \citep{Ginzburg_2016, ginzburg_2018}, and gas--poor formation \citep{lee22}. Similar to VE18, we find no planets within the gap using the stellar radii derived in this work, supporting the theory that previous ``gap planets'' are caused by uncertainties when constraining impact parameters without precise mean densities and short--cadence data \citep{petigura20}. 


\subsection{Gyrochronology}

Stellar ages are difficult to measure for low--mass stars, which are generally not amenable to age dating using isochrones or asteroseismology \citep{soderblom2010}. Gyrochronology is a powerful method to empirically determine stellar ages relying only on stellar rotation periods \citep[e.g.,][]{barnes_2007, Mamajek_2008, angus_2015, Angus2019, Spada2020}. Most \gyro relations have been calibrated on nearby, young open clusters, and therefore are limited for older stars \citep[e.g.,][]{Meibom2015}. Furthermore, the current sample of rotation periods in \kepler is biased since periods are measured for active stars that produce flux variations in the photometry, but stellar activity decreases for older stars \citep{Skumanich}. Calibration of these relations is required for confident age determination, but so far has produced discrepancies between clusters, \kepler stars, and nearby field stars \citep[e.g.,][]{angus_2015}.

In addition, the unexpected rapid rotation of stars more evolved than the Sun, a phenomenon known as weakened magnetic braking, has tested the accuracy of past age--rotation relations \citep[e.g.,][]{vansaders2016, Hall2021, Metcalfe2022, Bhalotia2024}. Fortunately, rotation rates as measured with rotational splitting of asteroseismic oscillation frequencies can help to calibrate age--rotation relation for older stars. It is therefore prudent to expand the sample of solar--like oscillators with high \snr for calculating age, and calibrating rotation--age relations for older stars.

Figure \ref{fig:prot_vs_numax} shows our sample with rotation periods and effective temperatures from \cite{berger_2020} against the complete sample from \cite{santos_2021} for reference. Rotation periods were obtained from \cite{santos_2021} (266 stars), \cite{McQuillan_2014} (52), \cite{garcia_2014} (91), and \cite{Hall2021} (37). The group of detections at \app5500 K above the \cite{santos_2021} distribution is because the asteroseismic sample is biased towards subgiants. Of the \withprot stars with rotation periods, 11 (not including targets from \cite{silva_2015} and \cite{silva_2017}) have sufficiently high data resolution -- as indicated by clear ridges in the \'echelle diagram -- to perform detailed modeling of oscillation modes for age determination \citep[e.g.,][]{Metcalfe2014, Creevey2017}.

\subsection{Comparison to Non--Detections}
To test the reliability of our catalog, we compared the sample of detections and non--detections against metrics that should correlate with the probability of finding oscillations in a star \citep[e.g.,][]{Chaplin2011_activity}. This includes the probability of detection (see Section 2.1.1 in \citet{Bhalotia2024}), white noise, and \kepler apparent magnitude (\kp). In Figure \ref{fig:snr_analysis}A, we show the white noise in the power spectrum (average power at frequencies $7000-8500$ \muhz) as a function of \kepler apparent magnitude for both detections and non--detections. We notice no detections fainter than \kp \app 14. Figure \ref{fig:snr_analysis}B shows the white noise as a function of length of time--series; while there are also many detections with short time--series, they are relatively less noisy than non--detections, with noise below \app 1000 ppm$^2/\mu$Hz. Figure \ref{fig:snr_analysis}C shows the white noise as a function of probability of detection. Detection probabilities were calculated based on the method by \citet{chaplin_2011}, using updated \gaia---derived stellar parameters from \citet{berger_2020}. Most stars (\app93\%) with detected oscillations have probability greater than 80\%. The 21 detections (\app3\% of the sample) with probability below 50\% are smaller stars ($R \lesssim 2$ \solrad, \logg $\gtrsim 4$ dex), and have an average noise level three times smaller than detections with probability above 50\%. 

We further investigated non--detections with an expected probability greater than 80\%, but found no common factor in stellar parameters for these stars. Stars with high probabilities that do not show detectable signal may have higher stellar activity, which is known to suppress oscillation amplitudes as investigated in Section \ref{sec:relation} \citep[e.g.,][]{Chaplin2011_activity}, but is not taken into account when calculating detection probability. Alternatively, noise in the light curve can also be a contributing factor.

\section{Conclusion}
We have created a homogeneous catalog of asteroseismic detections in \kepler short--cadence data. We use \pysyd to measure global asteroseismic quantities, and find solar--like oscillations in \total stars, of which \planets are planet hosts. We provide a homogeneous catalog with global asteroseismic properties such as \numax, \dnu, amplitude of oscillation, white noise, and height--to--background ratio. We also provide stellar radii and masses derived from scaling relations, and other stellar properties such as \kepler magnitude, effective temperature, \gaia DR3 RUWE, metallicity, and rotation period. Below is a summary of our main findings:

\begin{figure}[t!]
\begin{center}
\includegraphics[width=\linewidth,angle=0]{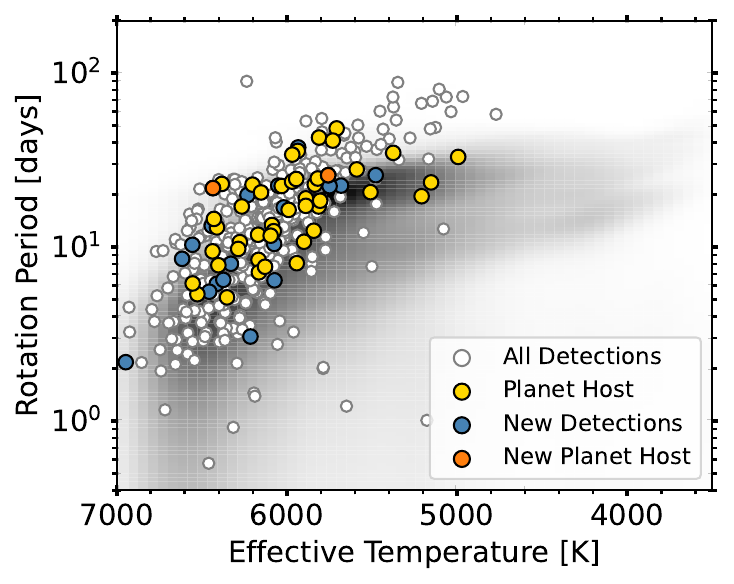}
\caption{Rotation periods as a function of effective temperatures for all detections (unfilled points), including new detections (blue), planet hosts (yellow), and new planet hosts (orange). The complete sample from \cite{santos_2021} is shown for reference as a density distribution.}
\label{fig:prot_vs_numax}
\end{center}
\end{figure}

\begin{enumerate}
    \setlength\itemsep{0em}
    \setlength\parskip{0em}
    \item We find \newdet new detections in the re--processed \kepler data, with a median uncertainty of 5.0\% and 2.7\% in \numax and \dnu, respectively. The new detections have no preferential location in the H--R diagram.
    \item For \total stars, we measure \numax and \dnu with median uncertainty of 3.4\% and 1.9\%, respectively. 
    We also derive seismic radii and masses with errors of 2.7\% and 10.5\%, respectively. The mean offset between our seismically derived radii and \gaia radii from \cite{berger_2020} is $0.8\pm0.2$\% with a scatter of 6\%. 
    
    \item We re--derive the amplitude scaling relation including a term for chromospheric activity, \rhk for main--sequence and subgiant stars (\teff $ \in [4900,6500]$ K, \logg $\in [3.5, 4.5]$ dex, and \rhk $\in [-5.4, -4.9]$). Compared to the scaling relation from \cite{Huber_2011}, the new scaling relation slightly reduces the offset between expected and observed oscillation amplitudes, and is statistically preferred when using a photometric temperature scale. The scatter in expected amplitude of $8-9\%$ as compared to the amplitude uncertainty of 11\% suggests that most, if not all, of the scatter in the expected amplitudes can be explained by measurement uncertainty. The revised scaling relation suggests that activity indicators can be used to predict oscillation amplitudes for future surveys.
    \item We find seven new detections that are \kepler planet hosts -- KOI 1032, KOI 302 or Kepler--516, KOI 126 or Kepler--469, KOI 94 or Kepler--89, KOI 684, KOI 76, and KOI 2533 or Kepler--1270 -- for which we provide global seismic parameters with detected oscillations. We derive planet radii using seismic stellar radii; the mean offset between our planet radii and those from \cite{vaneylen_2018} is $1.4\pm0.4\%$.
    \item In the final catalog, we include previously measured rotation periods for \withprot stars, where 11 have sufficiently high \snr for age determination with detailed modeling. We also provide global asteroseismic measurements for \planets planet hosts. 
\end{enumerate}

\begin{figure*}[t!]
\begin{center}
\includegraphics[width=\linewidth]{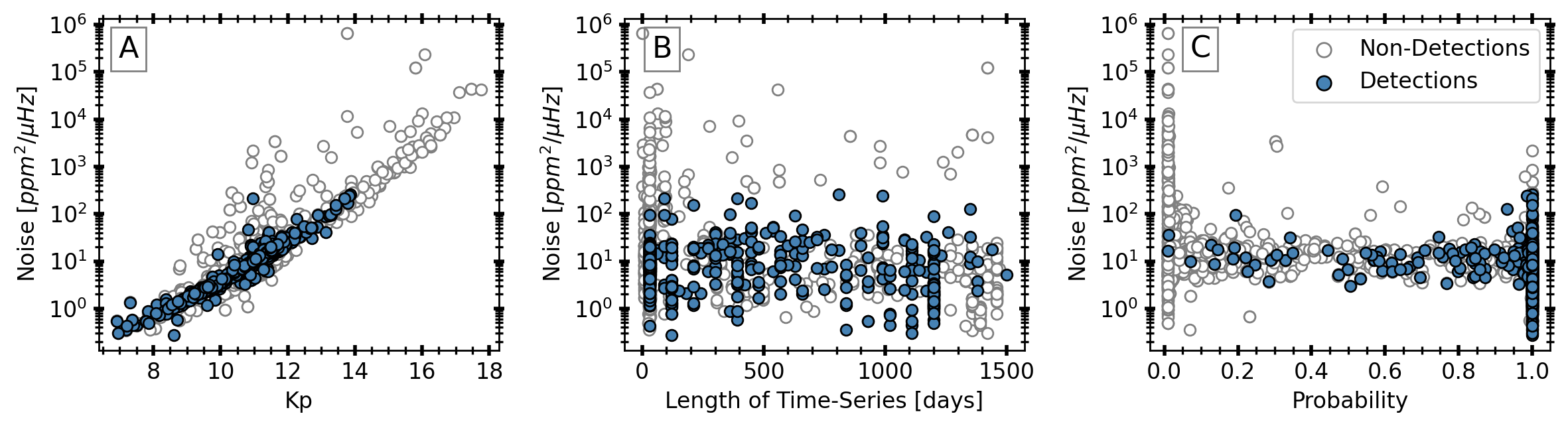}
\caption{\textit{Panel A:} Comparing distribution of white noise and \kepler apparent magnitude (\kp) between detections (blue points) and non--detections (unfilled). \textit{Panel B:} same as Panel A, but investigating white noise as a function of length of time--series. \textit{Panel C:}  same as Panel A, but investigating white noise as a function of detection probability.}
\label{fig:snr_analysis}
\end{center}
\end{figure*}

Our work presents the largest and most homogeneous catalog of solar--like oscillations in \kepler main--sequence and subgiant stars to date. Such catalogs are important for the development of data analysis tools, as benchmark stars for calibrating more indirect methods of determining stellar parameters \citep[e.g.][]{furlan18, sayeed_2021}, and for improving our understanding of stellar populations in the Galaxy \citep[e.g.,][]{chaplin11sci}. The \tess mission has already revealed hundreds of main--sequence and subgiant stars with solar--like oscillators \citep[e.g.,][]{chontos_2021, Huber2022, Hatt2023, metcalfe23, hon24, zhou24}. Insights and discoveries from \tess will shape science from the upcoming PLATO Mission \citep{plato2014, plato2024} and the Nancy Grace Roman Telescope \citep[][]{Spergel2015, Akeson2019}, which has a dedicated time--domain survey that is expected to yield \app$10^6$ asteroseismic detections in the Galactic center \citep{Huber2023,weiss25}. For example, investigation of the dependence of the oscillation amplitude scaling relation on stellar activity in different evolutionary phases and wavelength \citep{sreenivas25} will be important for the asteroseismic yield from upcoming missions.


\section*{Acknowledgments}
We are thankful to our anonymous referee for helpful comments that improved this manuscript. We thank Howard Isaacson for valuable input and comments on the paper. We gratefully acknowledge the tireless efforts of everyone involved with the \kepler mission. This work was supported by the National Aeronautics and Space Administration through the ADAP grant 80NSSC19K0597. M.S. is supported by the Research Corporation for Science Advancement through Scialog award  No. 26080. M.S. also thanks the LSSTC Data Science Fellowship Program, which is funded by LSSTC, NSF Cybertraining Grant No. 1829740, the Brinson Foundation, and the Moore Foundation; her participation in the program has benefited this work. D.H. acknowledges support from the Alfred P. Sloan Foundation and the Australian Research Council (FT200100871). This research made use of \lkurve, a Python package for \kepler and \tess data analysis.

Some of the data used in this paper were obtained at Keck Observatory, which is a private 501(c)3 non--profit organization operated as a scientific partnership among the California Institute of Technology, the University of California, and the National Aeronautics and Space Administration. The Observatory was made possible by the generous financial support of the W. M. Keck Foundation. The authors wish to recognize and acknowledge the very significant cultural role and reverence that the summit of Maunakea has always had within the Native Hawaiian community. We are most fortunate to have the opportunity to conduct observations from this mountain.

This research has made use of the Astrophysics Data System, funded by NASA under Cooperative Agreement 80NSSC21M00561. This research has made use of the SIMBAD database, CDS, Strasbourg Astronomical Observatory, France \citep{Wenger2000}, and the VizieR catalogue access tool, CDS, Strasbourg Astronomical Observatory, France \citep{vizier}. 

\software{astropy \citep{2013A&A...558A..33A,2018AJ....156..123A}, lightkurve \citep{lkurve}, Matplotlib \citep{matplotlib}, NumPy \citep{harris2020array}, 
Pandas \citep{mckinney-proc-scipy-2010}, 
\pysyd \citep{chontos_2022},  SciPy \citep{2020SciPy-NMeth}, TOPCAT \citep{Taylor2005_topcat} }

\facilities{Keck, \kepler}

\begin{deluxetable*}{ccc|cccc|cccc|ccc}[t!]
\tabletypesize{\scriptsize}
    \tablecolumns{14}
    \tablewidth{0pt}
    \tablecaption{Global asteroseismic values and errors measured with \pysyd for \total \kepler stars with detected solar--like oscillations. Oscillation amplitude at \numax, white noise, and height--to--background ratio are also provided. Previous values from other sources are provided for relevant targets. The source flag indicates the original source of asteroseismic measurements and the planet flag indicates whether the star is a planet host. \label{tb:pysyd-output} }
    \tablehead{
    \multicolumn{3}{c|}{} & \multicolumn{4}{c|}{Literature} & \multicolumn{7}{c}{pySYD} 
    \\ \cline{1-14}
    KIC ID & Source & Host &
    $\nu_{\text{max}}$ & $\sigma(\nu_{\text{max}})$ & \colhead{$\Delta \nu $} & \multicolumn{1}{c|}{$\sigma(\Delta \nu)$} & 
    $\nu_{\text{max}}$ & $\sigma(\nu_{\text{max}})$ & \colhead{$\Delta \nu $} &  \multicolumn{1}{c|}{$\sigma(\Delta \nu)$} & 
    \colhead{$A_\textrm{osc}$} & \colhead{Noise}  & \colhead{HBR}\\
    & & & 
    $[\mu \text{Hz}]$&  $[\mu \text{Hz}]$ &  $[\mu \text{Hz}]$ & $[\mu \text{Hz}]$ & 
    $[\mu \text{Hz}]$ &  $[\mu \text{Hz}]$ &  $[\mu \text{Hz}]$ &  $[\mu \text{Hz}]$ & [ppm$^2$/\muhz]& [ppm$^2$/\muhz]  &  }
    \startdata
    1430163 & 2 & 0 & 1867 & 92 & 84.60 & 2.00 & 1819 & 24 & 85.71 & 0.41 & 0.85 & 3.01 & 2.33 \\
    1435467 & 1 & 0 & 1382 & 19 & 70.56 & 0.09 & 1384 & 7 & 70.48 & 0.04 & 1.53 & 1.59 & 2.44 \\
    1725815 & 2 & 0 & 1045 & 47 & 55.40 & 1.30 & 1039 & 15 & 55.69 & 0.25 & 3.01 & 10.49 & 6.21 \\
    2010835 & 3 & 0 & 1312 & 19 & 72.73 & 4.52 & 1311 & 22 & 72.50 & 0.85 & 3.55 & 17.16 & 9.89 \\
    2162635 & 0 & 1 &  &  &  &  & 480 & 4 & 31.00 & 0.51 & 50.93 & 260.28 & 3.09 \\
    ... & ... & ... & ... & ... & ... & ... & ... & ... & ... & ... & ...
    \enddata
    \tablecomments{The planet flag is 1 if the star is a planet host, and 0 otherwise. The source for literature global asteroseismic values refer to the following catalogs -- 0: new detection, this work; 1: \cite{Serenelli_2017}; 2: \cite{Chaplin_2014}; 3: \cite{mathur_2021}; 4: \cite{Balona_2020}; 5: \cite{Huber_2013}; 6: \cite{Lundkvist2016}; 7: \cite{li_2020}; 8: \cite{Pinsonneault_2018}; 9: \cite{Lund_2017}; 10: \cite{mosser_2014}; 11: \cite{white_2012}; 12: \cite{Chontos_2019}; 13: \cite{Bhalotia2024}. The full table in machine--readable format can be found online.}
\end{deluxetable*}

\begin{deluxetable*}{ccccccccccccccc}[t!]
\tabletypesize{\scriptsize}
    \tablecolumns{14}
    \tablewidth{0pt}
    \tablecaption{Stellar properties of \total solar--like oscillators in our sample, including \teff from \cite{berger_2020}, RUWE from \gaia DR3, metallicity, and rotation periods. Stellar mass and radius were derived with scaling relations (Equations \ref{eq:sr_radius} \& \ref{eq:sr_mass}) using \pysyd measurements. \label{tb:misc} }
    \tablehead{
    KIC ID & \kp & \teff & $\sigma($\teff)& Radius &  $\sigma(R)$ & Mass & $\sigma(M)$ & RUWE & [Fe/H] & Source([Fe/H]) & Period & $\sigma($Period) & Source(P) \\
    & [mag] & [K] & [K] & [${\rm R_\odot}$] & [${\rm R_\odot}$] & [${\rm M_\odot}$] & [${\rm M_\odot}$] &  & [dex] & [dex] & [days] & [days] & --}
    \startdata
    1430163 & 9.58 & 6609 & 141 & 1.56 & 0.02 & 1.54 & 0.05 & 0.92 & --0.05 & 2 & 4.41 & 0.79 & 1 \\
    1435467 & 8.88 & 6391 & 128 & 1.73 & 0.01 & 1.41 & 0.03 & 5.73 & --0.03 & 1 & 6.59 & 0.66 & 1 \\
    1725815 & 10.83 & 6212 & 120 & 2.05 & 0.02 & 1.47 & 0.05 & 4.04 & --0.07 & 2 & 23.05 & 1.22 & 1 \\
    2010835 & 11.33 & 5927 & 117 & 1.49 & 0.03 & 0.96 & 0.07 & 0.97 &  &  &  &  &  \\
    2162635 & 13.86 & 4952 & 118 & 2.73 & 0.04 & 1.07 & 0.08 & 1.01 &  &  &  &  &  \\
    ... & ... & ... & ... & ... & ... & ... & ... & ... & ... & ... & ... &...
    \enddata
    \tablecomments{The source flag for [Fe/H] refers to the following three catalogs -- 1: \cite{Serenelli_2017}; 2: \cite{Buchhave_2015}; 3: \cite{Bruntt_2012}. The source flag for rotation period refers to the following three catalogs -- 1: \cite{santos_2021}; 2: \cite{McQuillan_2014}; 3: \cite{garcia_2014} The full table in machine--readable format can be found online.}
\end{deluxetable*}

\begin{deluxetable}{cccccc}[t!]
\tabletypesize{\scriptsize}
    \tablecolumns{6}
    \tablewidth{0pt}
    \tablecaption{KIC IDs and seismic measurements for the 22 stars which we did not confirm as detections. The source indicates the original paper that includes the target. \label{tb:unconfirmed}}
    \tablehead{
    KIC & Source & 
    $\Delta \nu $ & $\sigma(\Delta \nu)$  & $\nu_{\text{max}}$ & $\sigma(\nu_{\text{max}})$
    \\
    & & $[\mu \text{Hz}]$ & $[\mu \text{Hz}]$ & $[\mu \text{Hz}]$& $[\mu \text{Hz}]$ }
    \startdata
    3730801 & 1 & 74.5 & 2.0 &  &  \\
    11075448 & 1 & 78.1 & 2.9 &  &  \\
    7418476 & 3 & 64.8 & 1.9 & 1359 & 29 \\
    9109988 & 3 & 96.5 & 2.7 & 1876 & 11 \\
    10969935 & 3 & 29.5 & 2.7 & 462 & 8 \\
    6048403 & 4 & 53.6 & 0.1 & 1039.9 & 2.0 \\
    7833587 & 4 & 64.9 & 0.2 & 1302.9 & 3.0 \\
    10593626 & 5 & 137.50 &  1.40 &        &       \\
    6032981 & 5 &        &       &   35.1 &   0.6 \\
    4815520 & 6 & 136.0  & 0.3  &  &  \\
    5383248 & 6 & 149.4  & 0.5  &  &  \\
    6678383 & 6 & 56.5  & 0.4  &  &  \\
    7887791 & 6 & 156.1  & 0.6  &  &  \\
    7941200 & 6 & 130.1  & 0.5  &  &  \\
    8753657 & 6 & 121.0  & 0.4  &  &  \\
    9072639 & 6 & 85.2  & 0.3  &  &  \\
    9579641 & 6 & 131.2  & 0.3  &  &  \\
    10026544 & 6 & 26.0  & 0.6  &  &  \\
    10130039 & 6 & 143.5  & 0.4  &  &  \\
    10748390 & 6 & 186.6  & 0.4  &  &  \\
    11600889 & 6 & 131.8  & 0.5  &  &  \\
    11623629 & 6 & 161.1  & 0.4  &  &  \\
    \enddata
    \tablecomments{Source key -- 1: \cite{Serenelli_2017}; 3: \cite{mathur_2021}; 4: \cite{Balona_2020}; 5: \cite{Huber_2013}; 6: \cite{Lundkvist2016}.}
\end{deluxetable}

\begin{deluxetable*}{ccccccccccccccc}[t!]
\tabletypesize{\scriptsize}
    \tablecolumns{15}
    \tablewidth{0pt}
    \tablecaption{Spectroscopic stellar parameters for 127 stars with available spectra from \hires. We include \texttt{Specmatch} \teff, [Fe/H], and \logg with respective error bars. We also provide asteroseismic masses and radii calculated with both \texttt{Specmatch} \teff and isochrone based \teff. \label{tb:rphk}}
    \tablehead{
    KIC ID & $S_{\rm HK}$ & $B-V$ & \teff & $\rm \sigma(T_{eff})$ & [Fe/H] & $\rm \sigma([Fe/H])$ & \logg & $\rm \sigma(log \; \textit{g} )$ & $\log(\rm R'_{HK})$ &$\sigma(\log(\rm R'_{HK}))$& $R_{\rm spec}$&$M_{\rm spec}$&$R_{\rm iso}$& $M_{\rm iso}$ \\
     &  & & [K] & [K] & [dex] & [dex] & [dex] & [dex] &  & & [${\rm R_\odot}$] & [${\rm M_\odot}$] & [${\rm R_\odot}$] & [${\rm R_\odot}$]
    }
    \startdata
    1435467 & 0.14 & 0.48 & 6378 & 100 & 0.13 & 0.06 & 4.26 & 0.10 & --5.02 & 0.03 & 1.73 & 1.41 & 1.73 & 1.41 \\
    2852862 & 0.12 & 0.49 & 6252 & 100 & --0.04 & 0.06 & 4.10 & 0.10 & --5.23 & 0.04 & 2.11 & 1.48 & 2.13 & 1.53 \\
    3123191 & 0.10 & 0.48 & 6313 & 100 & --0.06 & 0.06 & 4.40 & 0.10 & --5.60 & 0.08 & 1.33 & 1.02 & 1.34 & 1.03 \\
    3424541 & 0.15 & 0.51 & 6267 & 100 & 0.24 & 0.06 & 3.87 & 0.10 & --4.99 & 0.03 & 2.80 & 2.04 & 2.82 & 2.09 \\
    3656476 & 0.14 & 0.68 & 5710 & 100 & 0.31 & 0.06 & 4.12 & 0.10 & --5.11 & 0.02 & 1.29 & 1.01 & 1.28 & 1.00 \\
    ... & ... & ... & ... & ... & ... & ... & ... & ... & ... & ... & ... & ... & ... & ... \\
    \enddata
    \tablecomments{The full table in machine--readable format is available online.}
\end{deluxetable*}

\bibliography{references}{}
\bibliographystyle{aasjournal}

\appendix
\renewcommand{\thefigure}{A\arabic{figure}} 
\setcounter{figure}{0}

\begin{figure*}[b!]
    \begin{center}
    \includegraphics[width=\linewidth,angle=0]{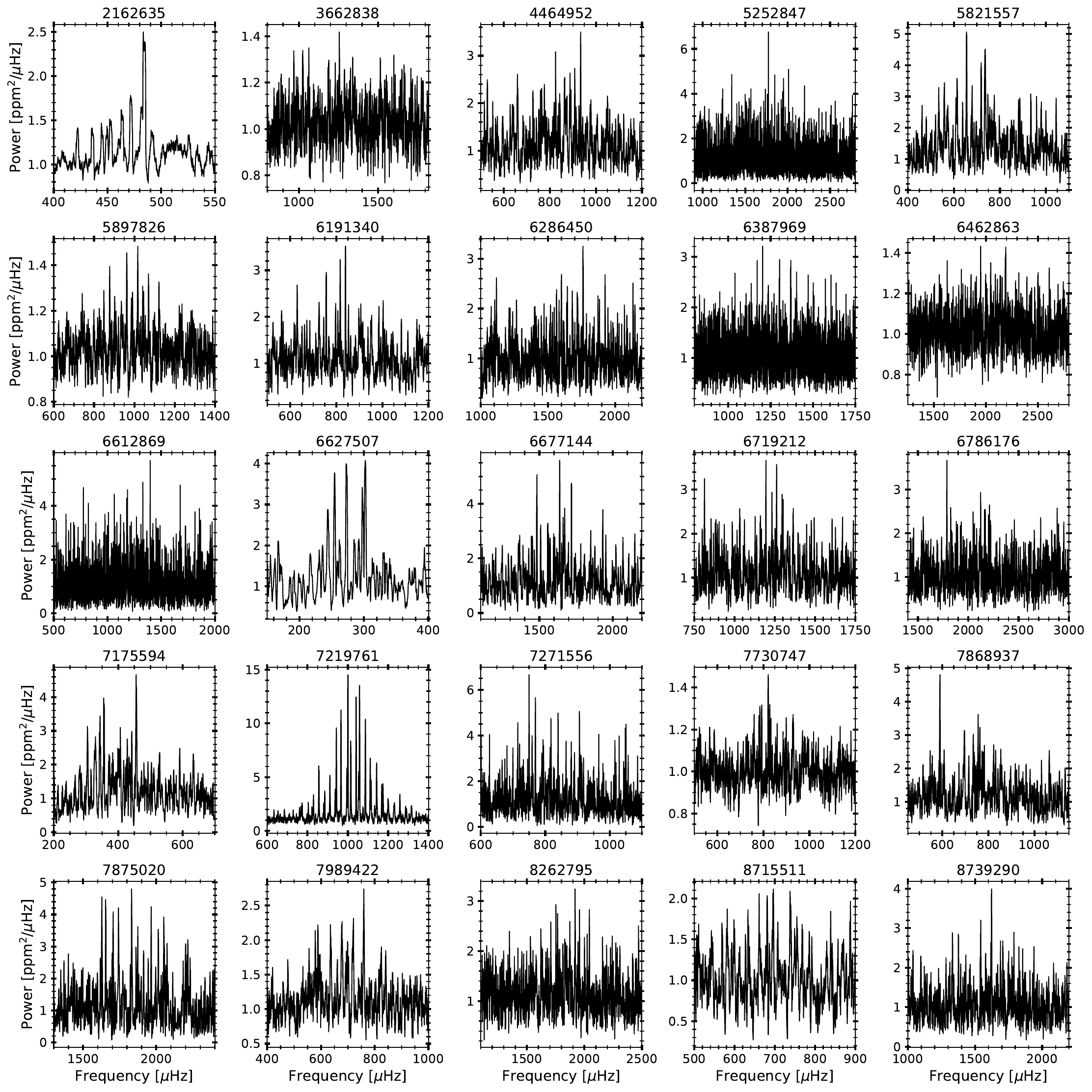}
    \caption{Smoothed and background corrected power spectra for 25 of the \newdet new detections as found by \pysyd, centered on \numax. }
\label{fig:new_dets_1}
\end{center}
\end{figure*}
\begin{figure*}
    \begin{center}
    \includegraphics[width=\linewidth,angle=0]{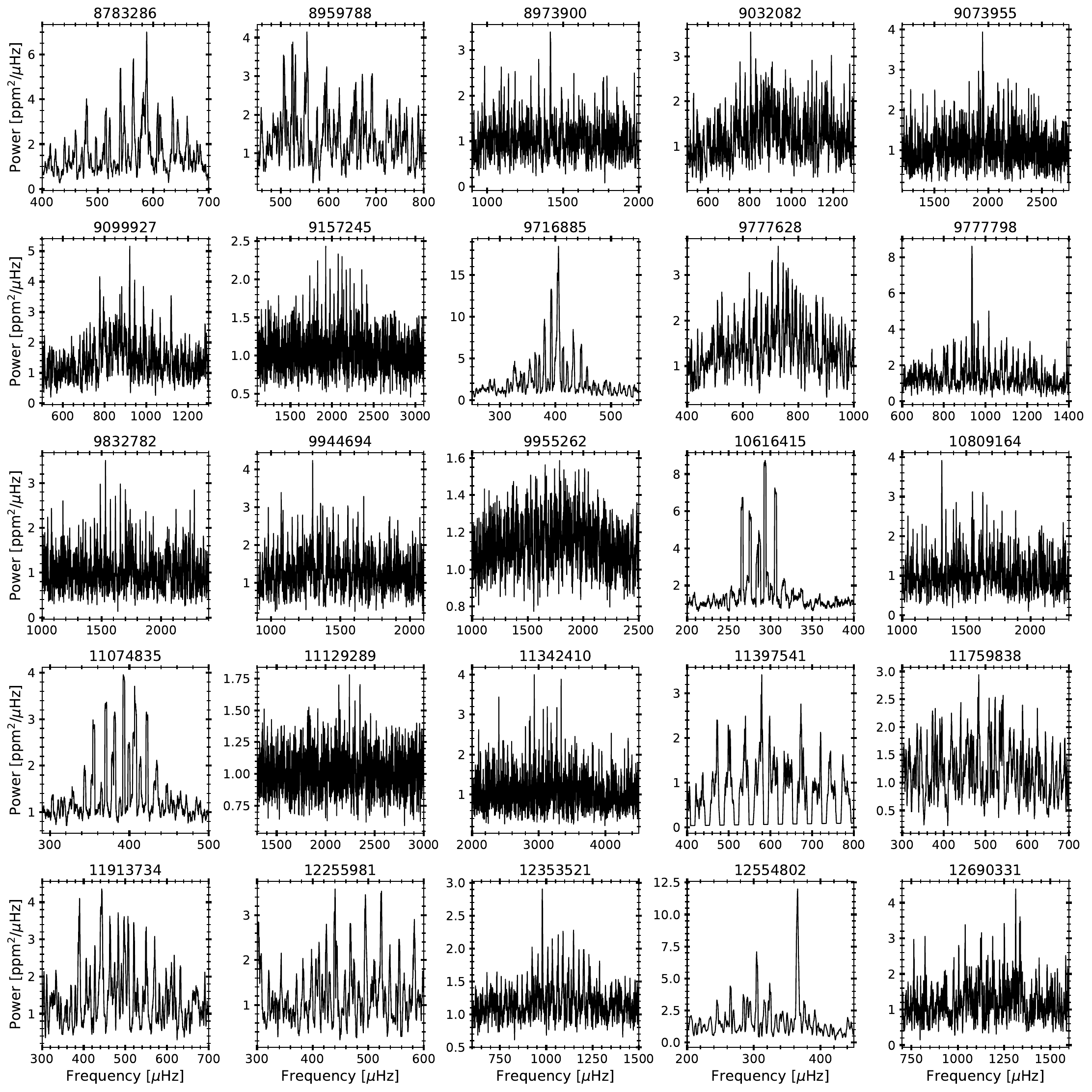}
    \caption{Same as Figure \ref{fig:new_dets_1}, for the other 25 of the \newdet new detections found in this work.}
\label{fig:new_dets_2}
\end{center}
\end{figure*}
\begin{figure*}[b!]
    \begin{center}
    \includegraphics[width=\linewidth,angle=0]{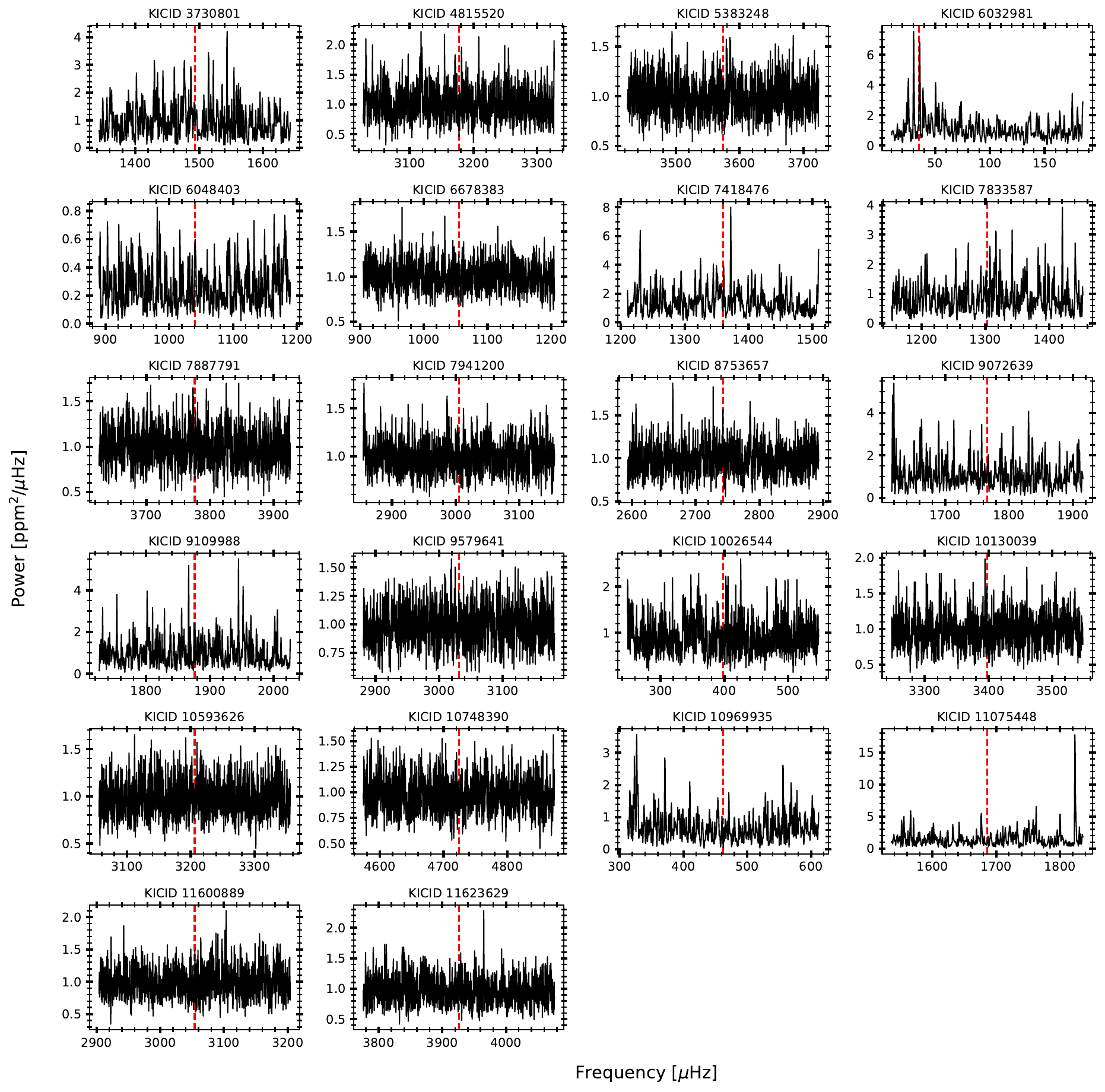}
    \caption{\kepler stars for which we did not find solar--like oscillations in DR25, but were previously classified as detections. Each subpanel shows the power spectrum of the target, centered on the expected \numax as indicated by the red dashed line.}
\label{fig:unconfirmed_dets}
\end{center}
\end{figure*}

Figures \ref{fig:new_dets_1} and \ref{fig:new_dets_2} show the power spectra of the \newdet new detections centered on \numax. In Figure \ref{fig:unconfirmed_dets}, we show the 22 stars that were previously classified as detections but we did not find solar--like oscillations in them.

\end{document}